\documentclass{elsart}

\usepackage{amssymb}
\usepackage{graphicx}
\usepackage{bm}
\usepackage{dsfont}
\usepackage{amsmath}
\usepackage{amscd}
\usepackage{float}

\begin{document}

\begin{frontmatter}



\title{Decays, contact $P$-wave interactions \\and hyperfine structure in $\Omega ^{-}$ exotic atoms $^{\star}$}
\thanks[label1]{Published in: Nucl. Phys. \textbf{A803} (2008) 173.}
\author{M. I. Krivoruchenko $^{ a,b}$}
\ead{mikhail.krivoruchenko@itep.ru}
\author{and Amand Faessler $^{ b}$}
\address{$^{a }$ Institute for Theoretical and Experimental Physics, B.
Cheremushkinskaya 25, 117259 Moscow, Russia}
\address{$^{b }$ Institut f\"{u}r Theoretische Physik, T\"{u}bingen
Universit\"{a}t, Auf der Morgenstelle 14, D-72076 T\"{u}bingen, Germany}

\begin{abstract}
Contact $P$-wave interactions connected to the Larmor interaction of a magnetic dipole and
Thomas spin precession in the filed of an electric quadrupole are described and their implications 
for spectroscopy of exotic $\Omega^{-}$-atoms are studied. In order to evaluate the magnitude of 
the contact $P$-wave interactions as compared to the conventional long-range interactions and the 
sensitivity of spectroscopic data to the $\Omega^{-}$-hyperon quadrupole moment, we consider $2P$ 
states of $\Omega ^{-}$ atoms formed with light stable nuclei with spins $I \geq 1/2$ and atomic numbers $Z \leq 10$.
The energy level splitting caused by the contact interactions is $2-5$ orders of magnitude smaller than the 
conventional long-range interactions. Strong decay widths of
$p\Omega ^{-}$ atoms due to reactions $p\Omega^{-} \rightarrow \Lambda \Xi^{0}$ and $p\Omega^{-} \rightarrow \Sigma \Xi$, 
induced by $t$-channel kaon exchanges, are calculated. $\Omega ^{-}$ atoms formed 
with the light nuclei have strong widths $5-6$ orders of magnitude higher than splitting caused by the contact interactions.
The low-$L$ pattern in the energy spectra of intermediate- and high-$Z$ $\Omega ^{-}$ atoms thus cannot be observed. 
The $\Omega ^{-}$ quadrupole moment can be measured by observing $X$-rays from circular 
transitions between high-$L$ levels in $\Omega^{-}$ exotic atoms. The effect of strong interactions 
in $^{208}$Pb$\Omega ^{-}$ atoms is negligible starting from $L \sim 10$.
The contact $P$-wave interactions exist in ordinary atoms and $\mu$-meson atoms.
\end{abstract}

\begin{keyword}
exotic atoms \sep $\Omega^{-}$-hyperons

\PACS 13.40.Em \sep 25.80.Nv \sep 32.10.Fn

\end{keyword}

\end{frontmatter}


\section{Introduction}
\setcounter{equation}{0}

In the early 70's, Goldhaber and Sternheimer \cite{GOLD} proposed to measure
the $\Omega ^{-}$-hyperon magnetic and quadrupole moments by detecting $X$
-rays from circular transitions of $\Omega ^{-}$-hyperons captured to atomic
orbits. Such method has been successful in extracting experimentally the magnetic 
moment of the $\Sigma^{-}$-hyperon from fine splitting in $\Sigma^{-}$ exotic atoms \cite{HERT}. 
The $\Omega ^{-}$ magnetic moment has been measured with high precision, however, 
by other techniques \cite{WALL,DIEH}. The measurement of the $\Omega^{-}$ electric quadrupole 
and magnetic octupole moments remains an open problem.

Among the decuplet baryons, the $\Omega ^{-}$-hyperon has weak decays only and a small width. It appears to be a suitable candidate for measurement of the static quadrupole moment. The measurement of the $\Omega ^{-}$-hyperon quadrupole moment would be helpful to understand better hadron structure and properties of quark interactions.

The $\Omega ^{-}$ exotic atoms are discussed in Refs. \cite{MOSK,RYND,GIAN,KRIV,BATT}. 
The $\Omega ^{-}$-hyperons are produced experimentally as relativistic particles. 
Stopping $\Omega ^{-}$ is a hard experimental task, since $\Omega ^{-}$-hyperons in 
matter dissolve to lighter hyperons by exchanging $K$-mesons with surrounding nuclei. 
The reaction $K^{-}p \to K^{-}K^{0}\Omega^{-}$ at threshold is in particular not well 
suited for producing slow $\Omega^{-}$-hyperons \cite{DOGA}. During the last three decades, 
there has been no progress in experimental studies of $\Omega ^{-}$ exotic atoms. 

Two events of stopped $\Xi^{-}$-hyperons in light emulsion nuclei at KEK have been interpreted 
as $\Xi^{-}$ atomic states bound with $^{12}$C \cite{AOKI93,AOKI95}. Future experiments for producing high rates
$\Xi$-hyperons at GSI are discussed in Refs. \cite{POCH,FERR}. Properties of $\Xi^{-}$ atoms are discussed in Refs. \cite{DOGA,BATT99}.

Recently Karl and Novikov \cite{KARL1,KARL2} made an interesting observation
on the existence of a contact $P$-wave interaction of two quadrupoles and
proposed to measure the $\Omega ^{-}$-hyperon quadrupole moment from the
hyperfine splitting of $P$-wave $\Omega ^{-}$ atomic states. The $\Omega ^{-}$-hyperon is
the only (almost) stable particle which can form bound states with a high-spin nucleus to exhibit
quadrupole-quadrupole interactions. 

The fine and hyperfine splittings in atoms relative to the ground state energy are of the order $(v/c)^2 \sim (\alpha Z)^2$, the hyperfine interaction is suppressed additionally by a factor $\sim m_{e}/M$, where $m_{e}$ is the electron mass and $M$ is the mass of nucleus, 
the Lamb shift is of the order $\sim \alpha (\alpha Z)^2 \log \frac{1}{\alpha Z}$, 
while the contact $P$-wave interaction is of the order 
$\alpha (\alpha Z)^3$. The additional smallness $\sim \alpha^2 Z$ as compared to the
dominant terms might be compensated in individual cases by large quadrupole moment 
of a high-$Z$ nucleus and/or a specific pattern of the quadrupole-quadrupole splitting.
In this work, we analyze hyperfine splitting in $\Omega ^{-}$ 
atoms by comparing numerically the magnitude of various interactions 
in $\Omega ^{-}$ atoms, formed with light stable nuclei, including spin-orbit interactions, 
spin-spin interaction, quadrupole-orbit interactions, which are of order $(v/c)^2$, and contact 
$P$-wave interactions of order $(v/c)^4$. \footnote{The quadrupole moments of nuclei increase with 
$Z$ roughly as $Z^{2/3}$, so the contact $P$-wave quadrupole-quadrupole interaction is well approximated as 
$(v/c)^4 \sim (\alpha Z)^4$.}

The measurement of energy splitting is possible provided widths of the corresponding energy levels 
are small. We calculate the strong decay widths of $p\Omega^{-}$ exotic atoms with arbitrary 
principal and orbital quantum numbers and give rough estimates of the strong decay widths of 
$\Omega^{-}$ exotic atoms formed with high-$Z$ nuclei.

The outline of the paper is as follows: In the next Sect., we discuss configuration mixing and exchange current 
contributions to quadrupole moments of the decuplet baryons and other static observables of baryons. In Sect. 3, 
a description of various interaction terms in bound systems, which appear in the nonrelativistic expansion of 
the one-photon exchange interaction potential between two high-spin particles, is given. The isotope dependence 
of the spin-orbit interaction is discussed. A contact $P$-wave electric quadrupole - magnetic dipole interaction 
is described and its magnitude is estimated and compared to other interactions in $\Omega^{-}$ atoms. In Sect. 4, 
we describe the calculation of strong decay widths of $p\Omega^{-}$ exotic atoms due to the processes $p\Omega^{-} \to \Lambda\Xi, \Sigma \Xi$. 
Strong decay widths of $2P$ states of $\Omega^{-}$ exotic atoms with light nuclei up to $^{19}$F are found to be up to three 
orders of magnitude higher than the dominant long-range interactions. Estimates made for circular transitions in 
$^{208}$Pb$\Omega^{-}$ exotic atoms give small strong decay widths starting from $L = n-1 \sim 10$. 

In Conclusion, we summarize the results.

\section{Configuration mixing vs two-body exchange currents} 
\setcounter{equation}{0}

Quark models are known to be very successful in the description of hadron properties. 
The one-gluon exchange describes the quadrupole moments of the decuplet baryons 
\cite{GERS,ISGU,RICH,QUAK,GIAN90,KRGI,WAGN00} and the non-vanishing neutron charge radius \cite{WAGN00,ELLI,NAIS,CLOS,CHRK,GIAN90}. 
In the framework of the Isgur-Karl nonrelativistic quark model, these quantities are simply related \cite{KRGI}:
\begin{equation}
Q_{\Delta^{+}} = \frac{2}{5} r_{n}^{2}\left|_{CM}\right.,
\label{OURS}
\end{equation}
where $Q_{\Delta^{+}}$ is the $\Delta^{+}$-isobar quadrupole moment and $r_{n}$ is the
neutron charge radius, determined by configuration mixing (CM)
in the baryon wave functions as illustrated on Fig. 1 (a). 

Let us discuss the status of CM effects in terms of the $v/c$ expansion. Spin-spin forces in the Fermi-Breit 
potential are of the order $V_{SS} \sim 1/(m^2r^3)$, where $m$ is the constituent quark mass. The corresponding 
perturbation of the baryon wave functions is of the order $\delta \Psi \sim \frac{V_{SS}}{\Delta E} \Psi \sim  \Psi/(m^2 \omega r^3)$, where $\Delta E \sim \omega$ ($\omega$ is the oscillator frequency). Thus the neutron charge radius and the 
quadrupole moments are of the order $1/(m^2 \omega r)$. The ratio between the neutron charge 
radius $r^2_n \sim 1/(m^2 \omega r)$ and the proton charge radius $r^2_p \sim 1/(m\omega)$
becomes  $r^2_n/r^2_p \sim \sqrt{m\omega}/m \sim v/c$, where we have used the relations $p^2 \sim m \omega$ for an oscillator 
and $p/m \sim v/c$. CM effects in the quadrupole moments and the neutron charge radius are therefore of the order $v/c$. Refs. \cite{GERS,ISGU,RICH,KRGI,GIAN90} and Refs. \cite{ELLI,NAIS,GIAN90} 
provide the calculations of $Q_{\Delta^{+}}$ and $r_{n}^{2}$, respectively, using the nonrelativistic quark model and Ref. \cite{CLOS} provides the calculation of $r_{n}^{2}$ using MIT bag model. Refs. \cite{GERS,ISGU,RICH,KRGI,ELLI,NAIS,CLOS,GIAN90} evaluate the CM. 
\footnote{A subclass of $Z$-diagrams shown on Fig. 2(d) of Ref. \cite{CLOS} 
vanishes. The remaining class of diagrams shown on Fig 2(b,c) corresponds to configuration mixing.}

Two-body exchange currents (EC) in bound systems contribute to observables also. They are associated to tree level $Z$-diagrams of the noncovariant perturbation theory, shown on Fig. 1 (b). 

\begin{figure}[!htb]
\vspace{3mm}
\begin{center}
\includegraphics[angle=0,width=7.0 cm]{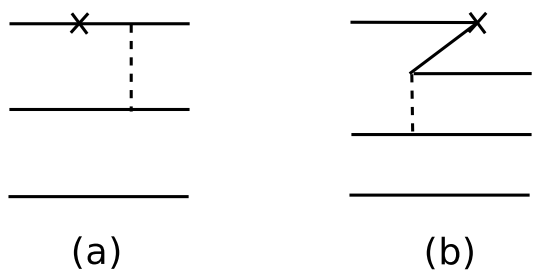}
\end{center}
\caption{Configuration mixing diagrams (a) and exchange current diagrams (b) contributing to an observable marked by the crosses. 
The solid lines are quarks, the dashed lines are gluons and mesons.}
\label{fig3}
\end{figure}

EC corrections to the charge density operator are of the order $1/(m^3 r^3)$ \cite{BUHE1,BUHE2}, so the corresponding corrections to the quadrupole moments and the neutron charge radius $\sim 1/(m^3 r)$. The ratio between EC correction to the neutron charge radius and the proton charge radius becomes $\sim \omega m /(m^3 r) \sim (v/c)^3$. EC corrections 
to the charge density operator and therefore to the quadrupole moments and the neutron charge radius are of the order $(v/c)^{3}$. In the framework of the nonrelativistic quark model, one can expect that EC effects are small as compared to CM effects for observables related to the charge density operator.

Precise measurements of the transition quadrupole moment $\Delta^{+} \to p\gamma$ give a value $Q_{\Delta^{+}p\gamma} = -0.108 \pm 0.009 \pm 0.034$ fm$^2$ \cite{QTRA} 
significantly higher than values predicted by the nonrelativistic quark models 
\cite{GERS,ISGU} based on evaluation of CM alone with  realistic quark core radii. One can expect that static quadrupole moments are undervalued too.
Buchmann, Hernandez and Faessler \cite{BUCH} conjectured that EC effects in observables related to the charge density operator are dominant. If one neglects CM effects and keep EC effects, one gets relation \cite{BUCH}
\begin{equation}
Q_{\Delta^{+}} = r_{n}^{2}\left|_{EC}\right.,
\end{equation}
which gives a higher value for the quadrupole moment of the $\Delta$.

Relativistic quark models sum up the $v/c$ series. It is thus instructive to compare the nonrelativistic quark model predictions with relativistic models. The experimental value of $Q_{\Delta^{+}p\gamma}$ appears to be three times higher than prediction of Ref. \cite{QUAK} based on the chiral bag model with account taken of CM and EC effects
\footnote{Ref. \cite{QUAK} treated gluon and quark fields classically. As shown in Ref. \cite{KOBZ}, 
summation of tree diagrams of the perturbation theory is equivalent to solving the classical equations of motion for gluon and quark fields, $Z$-diagrams are contained in the lower components of Dirac bispinors describing interacting quarks. 
Owing to quark self-interactions, results of Ref. \cite{QUAK} give quantum predictions.}.
The MIT bag model calculation of Close and Horgan \cite{CLOS} where CM effects are included only gives 
the neutron charge radius much smaller than that obtained in Ref. \cite{CHRK}. This result agrees qualitatively with the conjecture of Buchmann, Hernandez and Faessler \cite{BUCH} on the dominance of higher order $v/c$ terms in observables related to the charge density operator. In the chiral bag model, $Q_{\Delta^{+}p\gamma}$ and $r^2_n$ are still undervalued. EC corrections to $Q_{\Delta^{+}}$ and $r_{n}^{2}$ are calculated Refs.\cite{BUCH,WAGN00} using the nonrelativistic quark model and 
in Refs. \cite{QUAK,CHRK,KOBZ} using MIT and chiral bag models.

It is known that one-gluon exchange contributes to magnetic moments of baryons. EC contributions to the current density operator of nonrelativistic systems can be obtained from the Fermi-Breit potential by the minimal substitution $\mathbf{p} \to \mathbf{p} - e\mathbf{A}$ 
and taking derivative of the potential over $\mathbf{A}$. Magnetic moments of composite systems receive corrections $\delta \mu/\mu \sim 1/(mr) \sim v/c$.
The corresponding CM corrections due to the orthogonality of the space part of the quark 
wave functions are proportional to $\delta \mu/\mu \sim  (\frac{V_{SS}}{\Delta E})^2 \sim (v/c)^2$. 
In the framework of the nonrelativistic quark model, one can expect that EC corrections are large as compared to CM corrections when observables are related to the current density operator. Such a premise does not contradict to observations. CM corrections to baryon magnetic moments are calculated in Refs. \cite{ISGU80,WAGN00} in the nonrelativistic potential model and in Refs. \cite{KOBZ79,MIK84,MIK87,KOBZ,MIK90} using the MIT bag model.
EC corrections to baryon magnetic moments are calculated in Refs. \cite{MIK83,BUCH,WAGN00} using the nonrelativistic potential model and in Refs. \cite{MIK84,MIK87,KOBZ,MIK90} using MIT and chiral bag models.

Modern relativistic quark models \cite{Faessler:2006ky} treat exchange effects between quarks on the basis of
the covariant perturbation theory without explicit separation to CM and EC effects.

The mesons exchange effects are suppressed by the mass of the exchanged mesons. One can expect that in the $\Omega^{-}$-hyperon the role of mesons is less important. The measurement of the $\Omega^{-}$-hyperon quadrupole moment can be helpful to differentiate the gluon and meson exchange effects.

\section{Fine and hyperfine interactions} 
\setcounter{equation}{0}

Fine and hyperfine interactions in ordinary atoms are described in standard
textbooks (see e.g. \cite{LLQM,SOBE} and others), while specific features of $\Omega^{-}$ exotic atoms 
are discussed in Refs. \cite{MOSK,RYND,GIAN,KRIV,BATT}. The discussion is, however, 
restricted to spin-zero nuclei. For a high-spin nucleus, the pattern of energy 
levels is more complex due to the presence of higher nuclear multipole moments.

In this section, we summarize the known facts about interactions of high-spin particles, contributing to the energy 
level splitting to order $(v/c)^2$, and describe contact $P$-wave interactions of order $(v/c)^4$. 
Our purpose is to check the numerical magnitudes of various contributions to 
the energy splitting of an $\Omega^{-}$ atom in $L=1$ states with an intermediate mass nucleus. 
In Sect. 4, we compare the energy level splitting with widths of $\Omega^{-}$ exotic 
atom due to reactions $p\Omega^{-} \to \Lambda \Xi, \Sigma \Xi$.

\subsection{Isotope effect in spin-orbit interactions} 

The binding energy of electrons in atoms depends on the mass of nuclei, $M$, through 
the reduced electron mass 
\begin{equation}
m^{\prime} = \frac{mM}{m + M}. 
\end{equation}
In the hydrogen-like atoms,
\begin{equation}
E_{n} = - \frac{(\alpha Z)^2}{2n^2}m^{\prime},
\end{equation}
where $n$ is the principal quantum number. The isotope effect in the energy levels
of hydrogen-like atoms is measured experimentally and described in standard textbooks 
(see e.g. \cite{SOBE}).

The spin-orbit splitting in atoms represents a relativistic effect  
$\sim (v/c)^2$. In the hydrogen-like atoms,
\begin{equation}
\frac{v}{c} = \frac{\alpha Z}{n}.
\end{equation}
Corrections $\sim m_{e}/M \sim 5\cdot10^{-4}$ to the spin-orbit splitting 
in ordinary atoms are usually not discussed. 
However, exotic atoms such as antiproton and $\Sigma^{-}$-hyperon atoms are 
created and studied in the laboratory. 
In Ref. \cite{HERT}, the $\Sigma^{-}$ magnetic moment has been measured from 
fine structure in $\Sigma^{-}$ exotic atoms. 
In exotic atoms, the isotope effect becomes important.

The $LS$ potential consists of two parts. The Larmor part is connected to the interaction of 
the magnetic moment of the bound particle with the magnetic field generated by the nucleus 
in the co-moving frame of the bound particle. The second contribution is related to the Thomas 
spin precession. We thus write 
\begin{equation}
U_{LS} = U_{LS}^{L} + U_{LS}^{T}.
\end{equation}

The electrostatic potential created by a nucleus at rest with charge $-eZ$ has the form
\begin{equation}
\Phi = - \frac{eZ}{r},
\label{odin-1}
\end{equation}
where $e = - |e|$ is the electron charge and $r = |\mathbf{x}|$.
The electric field equals
\begin{equation}
\mathbf{E} = - \mathbf{\nabla}\Phi.
\end{equation} 
In the co-moving frame of the bound particle, the magnetic field can be found using the Lorentz transformation:
\begin{equation}
\mathbf{B}^{\prime} = - \mathbf{v} \times \mathbf{E} = - \frac{1}{m^{\prime}r}\frac{d\Phi}{dr}\mathbf{L},
\end{equation} 
where $\mathbf{v} = \mathbf{v}_1 - \mathbf{v}_2 = \mathbf{p}_1/m - \mathbf{p}_2/M = \mathbf{p}/m^{\prime}$ 
is the relative velocity, $\mathbf{p} \equiv \mathbf{p}_1 = - \mathbf{p}_2$ in the center-of-mass frame, 
$\mathbf{L} = \mathbf{x} \times \mathbf{p}$ where $\mathbf{x} = \mathbf{x}_1 - \mathbf{x}_2$. The indices 1 and 2 refer to bound particle and nucleus, respectively. 

The Larmor component of the spin-orbit interaction potential becomes
\begin{equation}
U_{LS}^{L} = - \frac{\mu}{S} \mathbf{S} \cdot \mathbf{B}^{\prime} 
= \frac{\alpha Z g}{2 m m^{\prime}r^3} \mathbf{L}\cdot \mathbf{S},
\end{equation} 
where 
\begin{equation}
\mu = \frac{eg}{2m}S 
\end{equation}
is the magnetic moment and $g$ is 
the gyromagnetic ratio of the particle (for electron $S = 1/2$ and $g = 2$). 

The angular frequency of Thomas precession, $\mathbf{\Omega}_T$, is 
related to the angular frequency of the orbital motion, $\mbox{\boldmath{$\omega$}}$:
\begin{equation}
\mathbf{\Omega}_T = \mbox{\boldmath{$\omega$}} (1 - \gamma),
\label{thomas}
\end{equation}
where $\gamma$ is the Lorentz factor of the moving particle. Equation (\ref{thomas}) is derived in Appendix A. 
The Hamiltonian producing the spin 
precession (\ref{thomas}) is given by
\begin{equation}
H_{T} = \mathbf{\Omega}_T \cdot \mathbf{S}.
\label{HAMTHOMAS}
\end{equation}
To the first order in $(v/c)^2$, one gets
\begin{equation}
U_{LS}^{T} = \mathbf{\Omega}_T \cdot \mathbf{S}
= - \frac{\alpha Z}{2m^2r^3}\mathbf{L}\cdot \mathbf{S}.
\label{HAMTH}
\end{equation}
Here, one used $\mathbf{L} = \mathbf{x}_1 \times \mathbf{p}_1 + \mathbf{x}_2 \times \mathbf{p}_2 
= (\mathbf{x}_1 - \mathbf{x}_2)\times \mathbf{p} = mrR\mbox{\boldmath{$\omega$}}$, 
where $R = m^{\prime}r/m$ is the distance from the center-of-mass of the system to the particle, $p = mR\omega$, and 
$mv^2/R = \alpha Z/r^2$ for particle on a circular orbit. In an external potential of scalar type with respect 
to the Lorentz group, the probing particle experiences the Thomas precession only, so its spin-orbit potential takes 
the form of Eq.(\ref{HAMTH}). The isotope dependence of scalar-exchange potentials, implied by Eq.(\ref{HAMTH}), is 
in agreement with Ref. \cite{GROM}.

The spin-orbit potential in the Coulomb field takes the form
\begin{equation}
U_{LS} = \frac{\alpha Z }{2 m r^3}\left(\frac{g}{m^{\prime}} - \frac{1}{m} \right) \mathbf{L}\cdot \mathbf{S}.
\label{LS}
\end{equation}
The Fermi-Breit potential for the $\Omega^{-}$-hyperon atom of Ref. \cite{GIAN}
contains Eq.(\ref{LS}). A distinct isotope dependence of the spin-orbit potential is used in Refs. \cite{GOLD,KARL2}. For $S =1/2$ and $g=2$ Eq.(\ref{LS}) is in agreement with Ref. \cite{LALI}, Chap. IX. 

For high-$Z$ atoms, the Dirac equation is usually used, modified to include the anomalous 
magnetic moment of the particle and with $m$ replaced by $m^{\prime}$. Borie \cite{BORI} developed efficient 
numerical schemes for calculation of energy eigenstates of relativistic 
atoms including the nucleus recoil corrections. The spin-orbit 
interaction obtained in Ref. \cite{BORI} by the non-relativistic reduction 
of the modified Dirac equation is in agreement with Eq.(\ref{LS}) to order 
$1/A$. This accuracy is sufficient for extracting the $\Sigma^{-}$ 
magnetic moment from the spin-orbit splitting of the high-$Z$ 
exotic atoms Pb-$\Sigma^{-}$ and W-$\Sigma^{-}$ \cite{HERT}. 
In low-$Z$ atoms such as $\bar{p}$-$^{3}$He or K$^{-}$-$p$ 
the $1/A^2$ corrections are important.

The magnetic field created by the orbital motion of the particle
acts on the magnetic moment of the nucleus. The nuclear spin $\mathbf{I}$ experiences the Thomas precession as well. 
The spin-orbit interaction potential has the form:
\begin{equation}
U_{IL} = \frac{\alpha Z}{2 M r^3}\left(\frac{g_{Z}}{m^{\prime}} - \frac{1}{M} \right) \mathbf{I} \cdot \mathbf{L}.
\label{IL}
\end{equation}
The magnetic moment of the nucleus is defined by
\begin{equation}
\mu_{Z} = -\frac{eZg_{Z}}{2M}I.
\end{equation}
For high-$Z$ nuclei, the Larmor contribution to $U_{IL}$ is of order of unity, whereas 
the Thomas precession is suppressed as $1/A$. In the potential $U_{LS}$ these contributions 
are of the same order in $1/A$.

\subsection{Spin-spin interaction}

The long-range interaction of two magnetic dipoles is well known (see e.g. \cite{LLQM}, Chap. XVI): 
\begin{equation}
U_{IS} = \frac{3 \alpha Z g_{Z} g}{16mMr^3} \tau^{\alpha \beta}(\mathbf{n},\mathbf{n}) \tau^{\alpha \beta}(\mathbf{I},\mathbf{S}).
\label{IS}
\end{equation}
The irreducible tensor $\tau^{\alpha \beta}(\mathbf{a},\mathbf{b})$ with space indices 
$\alpha, \beta = 1,2,3$ is defined by
\begin{equation}
\tau^{\alpha \beta}(\mathbf{a},\mathbf{b}) = a^{\alpha} b^{\beta} + a^{\beta} b^{\alpha} 
- \frac{2}{3} \mathbf{a \cdot b} \delta^{\alpha \beta}. \label{IRREBIS}
\end{equation}
Its properties are described in Appendix B.

\subsection{Quadrupole-orbit interactions}

The electrostatic potential created by a spin-$I$ nucleus gives rise to
hyperfine splitting connected to the interaction of the nucleus electric quadrupole 
moment with electrostatic field created by the orbital motion of the bound particle. 
In terms of the nucleus spin $\mathbf{I}$, the nucleus electric quadrupole moment has 
the form 
\begin{equation}
Q_{Z}^{\alpha \beta }=-\frac{3eQ_{Z}}{2I(2I-1)}\tau ^{\alpha \beta }(\mathbf{%
I},\mathbf{I}).
\end{equation}
It is normalized by
\begin{equation}
<II|Q_{Z}^{zz}|II>=-eQ_{Z}.
\end{equation}
For nuclear electric quadrupole moments, experiments quote $Q_{Z}$ i.e. the proton charge, $-e$, is usually omitted.

The quadrupole-orbit interaction has the form (see e.g. \cite {SOBE}) 
\begin{equation}
U_{Q_{Z}L}=-\frac{\alpha }{4}\frac{3Q_{Z}}{2I(2I-1)}\frac{1}{r^{3}}%
\tau ^{\alpha \beta }(\mathbf{n},\mathbf{n})\tau ^{\alpha \beta }(\mathbf{I},%
\mathbf{I}).  
\label{QL}
\end{equation}
The hyperfine splitting caused by the potential $U_{Q_{Z}L}$ is used to measure the electric
quadrupole moments of nuclei from spectroscopy of 
ordinary atoms \cite{SOBE} and $X$-ray spectroscopy of $\pi$-, $K$,- and 
$\mu$-meson exotic atoms (see e.g. \cite{Mu,Pi-Mu,Ka-Pi}).

The nucleus Coulomb field interacts with the quadrupole moment of 
$\Omega^{-}$
\begin{equation}
Q^{\alpha \beta }=\frac{3eQ}{2S(2S-1)}\tau ^{\alpha \beta }(\mathbf{S%
},\mathbf{S}).
\end{equation} 
It is normalized as follows:
\begin{equation}
<SS|Q^{zz}|SS>=eQ.
\end{equation} 

The interaction potential has the form of Eq.(\ref{QL}) with 
the replacements $\alpha \leftrightarrow \alpha Z$, 
$Q_{Z}\leftrightarrow Q$, and $\mathbf{I} \leftrightarrow \mathbf{S}$:
\begin{equation}
U_{LQ} = - \frac{\alpha Z}{4}\frac{3Q}{2S(2S - 1)}\frac{1}{r^3}
\tau^{\alpha \beta}(\mathbf{n},\mathbf{n})\tau^{\alpha \beta}(\mathbf{S},\mathbf{S}).
\end{equation}

\subsection{Quadrupole-spin contact $P$-wave interactions}

Contact $S$-wave interactions of baryons originating from meson exchanges are usually 
omitted from the start, since those interactions are made to vanish by the repulsive 
core anyway. The $S$-wave contact terms, generated by photon exchanges, can be set equal 
to zero in the baryon-baryon potentials also.

Contact $P$-wave interactions depend on the gradient 
of wave function at the origin. Assuming the wave function is 
suppressed as $\Psi(0) \sim \exp(-\sqrt{2mU_{0}}b) \ll 1$ where $U_{0}$ is height of the core and 
$b$ is the core radius, one finds that $\Psi(0)^{\prime}$ is suppressed as 
$\Psi(0)^{\prime} \sim \sqrt{2mU_{0}}a_B \exp(-\sqrt{2mU_{0}}b)$, 
where $a_B$ is the Bohr radius, $a_B \gg b$. One sees that suppression of the 
contact $P$-wave interactions is less effective and can in principle be compensated by the large 
factor $\sqrt{2mU_{0}}a_B$.

We analyse contact $P$-wave interactions quantitatively and, as a first 
approximation, assuming no effect from the repulsive core exists at all.

Let us consider the quadrupole part of the electrostatic field
\begin{equation}
E^{\gamma} = - \nabla^{\gamma}\Phi = - \frac{1}{6}Q_{Z}^{\alpha \beta}\nabla^{\gamma}\nabla^{\alpha}\nabla^{\beta}\frac{1}{r}.
\label{EEEE}
\end{equation} 
Tensor $\nabla^{\gamma}\nabla^{\alpha}\nabla^{\beta}\frac{1}{r}$ entering this expression can be split 
into two parts with angular momenta $L = 3$ and $L = 1$:
\begin{equation}
\nabla^{\alpha}\nabla^{\beta}\nabla^{\gamma} \frac{1}{r} 
= T_{[3]}^{\alpha \beta \gamma} + T_{[1]}^{\alpha \beta \gamma},
\end{equation}
where
\begin{equation}
T_{[1]}^{\alpha \beta \gamma} = \frac{1}{5}(\delta^{\alpha \beta} \nabla^{\gamma} + \delta^{\beta \gamma } \nabla^{\alpha } + \delta^{\gamma \alpha } \nabla^{\beta}) \triangle \frac{1}{r} 
= - \frac{4\pi}{5}(\delta^{\alpha \beta} \nabla^{\gamma} + \delta^{\beta \gamma } \nabla^{\alpha } + \delta^{\gamma \alpha } \nabla^{\beta}) \delta (\mathbf{x}). \nonumber
\end{equation}

In the co-moving coordinate system of the $\Omega^{-}$ hyperon the induced magnetic field has the form
\begin{equation}
{B}^{\prime \lambda} = - \epsilon^{\lambda \kappa \gamma} v^{\kappa} E^{\gamma}
\end{equation}
where $v^{\kappa} = p^{\kappa}/m^{\prime}$ is velocity, $p^{\kappa}$ momentum and $m^{\prime}$ reduced mass of $\Omega^{-}$, $\epsilon^{\lambda \kappa \gamma}$ is the totally antisymmetric tensor such that $\epsilon^{123} = 1$. 
The Larmor interaction energy of the $\Omega^{-}$ magnetic moment with the magnetic field is given by
\begin{equation}
U_{Q_{Z}S}^{L} = - \frac{\mu S^{\lambda}}{S} {B}^{\prime \lambda}.
\label{KJHU}
\end{equation}
The contact part of the interaction has the form
\begin{equation}
U_{Q_{Z}S}^{cL} = \frac{2\pi \mu }{15 S m^{\prime}} \epsilon^{\lambda \gamma \beta} 
Q_{Z}^{\alpha \beta} S^{\lambda} (p^{\gamma} (\nabla^{\alpha} \delta(\mathbf{x})) + (\nabla^{\alpha} \delta(\mathbf{x})) p^{\gamma}).
\label{POI}
\end{equation}
The term entering the brackets, being averaged over the $L=1$ state, gives
\begin{eqnarray}
\int d\mathbf{x} Y^{*}_{1m'}(\mathbf{n})R_{n1}(r)
\left(p^{\alpha} (\nabla^{\beta} \delta(\mathbf{x})) + (\nabla^{\beta} \delta(\mathbf{x})) p^{\alpha}\right) 
Y_{1m}(\mathbf{n})R_{n1}(r) \nonumber \\
= \frac{3}{4\pi} \epsilon^{\alpha \beta \gamma} <1m'|L^{\gamma}|1m>R^{\prime 2}_{n1}(0),
\nonumber
\end{eqnarray}
where $R^{\prime}_{nL}(0)$ is the derivative of the radial wave function at the origin Eq.(\ref{R0}). 
We thus obtain
\begin{equation}
U_{Q_{Z}S}^{cL} = \frac{\alpha g }{40 m m^{\prime}} \frac{3Q_{Z}}{2I(2I - 1)} 
\tau^{\alpha \beta}(\mathbf{I},\mathbf{I}) \tau^{\alpha \beta}(\mathbf{S},\mathbf{L})R^{\prime 2}_{n1}(0).
\label{QS}
\end{equation}

For evaluation of the Thomas precession component of the interaction energy, we determine from Eq.(A.22) 
the Thomas precession frequency,
\begin{equation}
\mathbf{\Omega}_{T} \approx - \frac{1 }{2} \mathbf{\mathbf{v}}_{1} \times \dot{\mathbf{v}}_{1},
\label{FREQ}
\end{equation}
and use Eq.(\ref{HAMTHOMAS}). For circular motion $\mathbf{\Omega}_{T}$ is in agreement with the nonrelativistic limit of Eq.(\ref{thomas}). Taking for $\mathbf{E}$ expression (\ref{EEEE}) and substituting its contact part into the equation of motion $m{d}\mathbf{v}_{1}/{dt} = e\mathbf{E}$, and 
further, ${d}\mathbf{v}_{1}/{dt}$ into Eqs.(\ref{FREQ}), we obtain the Thomas component, $U_{Q_{Z}S}^{cT}$, of the contact interaction potential. If we would use for $\mathbf{E}$ expression (III.6), 
we could get $U_{LS}^{T}$.

The sum of the Larmor and Thomas interactions takes the form
\begin{equation}
U_{Q_{Z}S}^{c} = \frac{\alpha }{40 m} \left(\frac{g}{ m^{\prime}} - \frac{1}{m}\right)\frac{3Q_{Z}}{2I(2I - 1)} 
\tau^{\alpha \beta}(\mathbf{I},\mathbf{I}) \tau^{\alpha \beta}(\mathbf{S},\mathbf{L})R^{\prime 2}_{n1}(0).
\label{QSFULL}
\end{equation}
The factor ${g}/{ m^{\prime}} - {1}/{m}$ appears both in the spin-orbit and quadrupole-spin interactions. It gives the 
one-half reduction of the energy levels splitting for Dirac spin-1/2 particles in heavy nuclei.

Similarly, the interaction potential of the nucleus spin and the $\Omega^{-}$ quadrupole moment can be found to be
\begin{equation}
U_{IQ}^{c} = \frac{\alpha Z}{40 M} \left(\frac{g_{Z}}{ m^{\prime}} - \frac{1}{M}\right) \frac{3Q}{2S(2S - 1)} 
\tau^{\alpha \beta}(\mathbf{S},\mathbf{S}) \tau^{\alpha \beta}(\mathbf{I},\mathbf{L})R^{\prime 2}_{n1}(0). 
\label{IQ}
\end{equation}
A modification of the contact interactions due to relativistic effects and finite volume of the nuclei is discussed 
in Appendix C, with applications to the muonic atom. It is shown, in particular, that the $L=3$ component of the electrostatic potential and the Darwin term generate contact $P$-wave interactions also.

\begin{table}[]
\addtolength{\tabcolsep}{-3pt}
\centering
\renewcommand{\arraystretch}{1.1}
\caption{
The magnitudes of long-range and contact $P$-wave interactions and strong decay widths of $\Omega^{-}$ atoms in $2P$ states,
formed with light stable spin $I \geq \frac{1}{2}$ nuclei. The experimental values of the nuclear dipole magnetic moments 
$\mu_{Z}$ and the nuclear electric quadrupole moments $Q_{Z}$ are taken from Ref. \cite{RAGH}, errors are not displayed.
The multipole moments of $\Omega^{-}$: $\mu =-2.02$ n.m. \cite{WALL,DIEH}, $Q=-2.8\times 10^{-2}$ fm$^{2}$ \cite{KRGI}. 
$U^{[2]}_{\max}$ is the maximum value over $F$ of the root mean square of eigenvalues of the matrix elements $<FJ|U^{[2]}|FJ^{\prime}>$, 
$U^{[4]}_{\max}$ is defined similarly. For $^{1}$H, the strong decay width is given by Eq.(\ref{WIDTH}),
in other cases $\Gamma$ are estimates based on Eq.(\ref{APPR}).
}
\vspace{2mm}
\label{tab:table5}
\begin{tabular}{|c|ccccccc|}
\hline
 Nuclei                   & $^{1}$H           & $^{2}$H           & $^{3}$H           & $^{3}$He          & $^{6}$Li          & $^{7}$Li          & $^{9}$Be    \\ \hline \hline
$I$                       & 1/2               & 1                 & 1/2               & 1/2               & 1                 & 3/2               & 3/2                \\ \hline
$\mu_{Z}$ [n.m.]          & 2.79              & 0.86              & 2.98              & -2.13             & 0.82              & 3.26              & -1.18              \\ \hline
$Q_{Z}$ [fm$^2$]          &                   & 0.29              &                   &                   & -0.08             & -4.06             & 5.3                \\ \hline
$U^{[2]}_{\max}$ [keV]    & $9\times 10^{-6}$ & $3\times 10^{-4}$ & $10^{-4}$         & $6\times 10^{-3}$ & $2\times 10^{-2}$ & 0.2               & 1                  \\ \hline
$U^{[4]}_{\max}$ [keV]    & $4\times 10^{-10}$& $7\times 10^{-9}$ & $4\times 10^{-9}$ & $10^{-7}$         & $2\times 10^{-6}$ & $3\times 10^{-5}$ & $3 \times 10^{-4}$ \\ \hline
$\Gamma        $ [keV]    & $6.3\times 10^{-6}$ &                   &                   &                   & 10                & 20                & 100         \\ \hline \hline
 Nuclei                   & $^{10}$B          & $^{11}$B          & $^{13}$C          & $^{14}$N          & $^{15}$N          & $^{17}$O          & $^{19}$F    \\ \hline \hline
 $I$                      & 3                 & 3/2               & 1/2               & 1                 & 1/2               & 5/2               & 1/2                \\ \hline
 $\mu_{Z}$    [n.m.]      & 1.80              & 2.69              & 0.70              & 0.40              & -0.28             & -1.89             & 2.63               \\ \hline
 $Q_{Z}$  [fm$^2$]        & 8.47              & 4.07              &                   & 2.00              &                   & -2.58             &                    \\ \hline
$U^{[2]}_{\max}$ [keV]    & 2                 & 2                 &     0.5           & 3                 &    1              & 3                 &    2               \\ \hline
$U^{[4]}_{\max}$ [keV]    & $4\times 10^{-3}$ & $8\times 10^{-4}$ & $3\times 10^{-5}$ & $6\times 10^{-3}$ & $3\times 10^{-5}$ & $2\times 10^{-2}$ & $10^{-3}$          \\ \hline
$\Gamma      $ [keV]      & 350               & 500               & $10^{ 3}$         & $2\times 10^{3}$  & $4\times 10^{3}$  & $7\times 10^{3}$  & $10^{4}$   \\
\hline
\end{tabular}
\end{table}


\subsection{Quadrupole-quadrupole contact $P$-wave interaction}

The contact part of the quadrupole-quadrupole interaction proposed in Ref. \cite{KARL1} 
looks like
\begin{equation}
U_{Q_{Z}Q}^{c} = - \frac{4\pi}{63} Q_{Z}^{\alpha \gamma} Q^{\beta \gamma} 
(\nabla^{\alpha} \nabla^{\beta} - \frac{1}{10} \delta^{\alpha \beta} \triangle)\delta(\mathbf{x}).
\end{equation}
After integration over the angles and some additional algebra, one gets for $L=1$ multiplet 
\begin{eqnarray}
U_{Q_{Z}Q}^{c} &=& \frac{\alpha}{63} \frac{3Q_{Z}}{2I(2I - 1)} \frac{3Q}{2S(2S - 1)} \label{QQ} \\
&\times& \left(
\frac{7}{5} \tau^{\alpha \beta}(\mathbf{I},\mathbf{I})\tau^{\alpha \beta}(\mathbf{S},\mathbf{S}) 
- 3 \tau^{\gamma \alpha}(\mathbf{I},\mathbf{I}) \tau^{\alpha \beta}(\mathbf{S},\mathbf{S})\tau^{\beta \gamma}(\mathbf{L},\mathbf{L})
\right)R^{\prime 2}_{n1}(0).
\nonumber
\end{eqnarray}

\subsection{Numerical estimates of long-range and contact $P$-wave interactions in $\Omega^{-}$ exotic atoms}

The contact $P$-wave interactions and the long-range interactions in $^{14}$N$\Omega ^{-}$ exotic 
atom are compared in Appendix B. 
In Table \ref{tab:table5}, we report the magnitudes of the interaction energies 
\begin{equation}
U^{[2]} = U_{LS} + U_{IL} + U_{IS} + U_{Q_{Z}L} + U_{LQ} 
\label{U2}
\end{equation} 
and 
\begin{equation}
U^{[4]} = U_{IQ}^{cL} + U_{Q_{Z}S}^{cL} + U_{Q_{Z}Q}^{c}
\label{U4}
\end{equation} 
for light nuclei with atomic numbers below 10 and spins $I \geq 1/2$. The quantities $U^{[2r]}_{\max}$ given in 
Table \ref{tab:table5} for $r=1,2$ are defined by
\begin{equation}
U^{[2r]}_{\max} = \max_{F}\sqrt{\frac{\mathrm{Tr}[(U^{[2r]})^{2}]}{\mathrm{Tr} [1]}}.
\end{equation} 
$U^{[2r]}_{\max}$ is the maximum over $F$-multiplets of the root mean square of eigenvalues 
of $U^{[2r]}$. $U^{[2]}_{\max}$ and $U^{[4]}_{\max}$
give typical strengths of the $(v/c)^2$ and $(v/c)^4$ interactions. We observe that $U^{[2]}_{\max}$ is 2-5 orders of magnitude greater than $U^{[4]}_{\max}$.
 
One can see that the variations of $<FJ^{\prime}|U^{[4]}|FJ>$ in different nuclei are irregular and high. 
Counting of powers of $v/c \sim \alpha Z/n$ 
gives however the right first idea on the magnitudes of various interactions.

\section{Decays of $\Omega^{-}$ exotic atoms} 
\setcounter{equation}{0}

Decays of the $\Omega ^{-}$ exotic atoms proceed due to kaon $t$-channel
exchanges between nucleons of the nucleus and the $\Omega ^{-}$-hyperon: 
$p\Omega ^{-}\rightarrow \Lambda \Xi ^{0}$ + 180 MeV or $\Sigma \Xi $ + 100
MeV. These channels are shown on Fig. 2. 

\subsection{Transition vertices: Relativistic expressions}

The effective vertices of the transitions $p \to Y K^{+},Y K^{*+}$ where $Y = \Lambda, \Sigma^{0}$ and 
$\Omega ^{-} K^{+} \to \Xi ^{0}$, $\Omega ^{-} K^{*+} \to \Xi ^{0}$ may be written in the form
\begin{eqnarray}
<Y|J_{P}(0)|N> &=&g_{NYK}\bar{u}(p_{Y },s_{Y })i\gamma _{5}u(p_{N},s_{N}),  \label{EL1} \\
<\Xi |J_{P}(0)|\Omega> &=& - \frac{g_{\Omega \Xi K}}{m_{\Xi }}\bar{u}(p_{\Xi },s_{\Xi })q^{\mu }u_{\mu }(p_{\Omega },s_{\Omega }), 
\label{EL2} \\
<Y|J^{\mu}_{V}(0)|N> &=& \bar{u}(p_{Y },s_{Y})(g_{NY K^{*}}\gamma ^{\mu }+\frac{f_{NY K^{*}}}{2m_{N}}i\sigma^{\mu \nu }q_{\nu }) u(p_{N},s_{N}), \label{EL3} \\
<\Xi |J^{\mu}_{V}(0)|\Omega> &=& - i\sum_{i} \frac{f_{\Omega \Xi K^{*}}^{[i]}}{m_{\Xi}^{2}}  \bar{u}(p_{\Xi },s_{\Xi })\bar{\Gamma}_{i}^{ \nu \mu}u_{\nu }(p_{\Omega },s_{\Omega }). \label{EL4}
\end{eqnarray}
Here, $J_{P}(x)$ is the pseudoscalar current coupled to the pseudoscalar mesons, $J^{\mu}_{V}(x)$ is the vector current 
coupled to the vector mesons, and
\begin{eqnarray}
\bar{\Gamma}_{1 \nu \mu} &=& m_{\Omega}(q_{\nu }\gamma _{\mu }-\hat{q}%
g_{\nu \mu })\gamma _{5}, \\
\bar{\Gamma}_{2 \nu \mu} &=& - (q_{\nu }P_{\mu}-q\cdot Pg_{\nu \mu
})\gamma _{5}, \\
\bar{\Gamma}_{3 \nu \mu} &=& - (q_{\nu }q_{\mu }-q^{2}g_{\nu \mu})\gamma
_{5}.
\end{eqnarray}
The vertices of the transitions $p \to \Sigma^{+} K^{0}, \Sigma^{+} K^{*0}$ and 
$\Omega ^{-} K^{0} \to \Xi ^{-}$, $\Omega ^{-} K^{*0} \to \Xi ^{-}$ are related by isotopic symmetry 
with (\ref{EL1}) - (\ref{EL4}). The amplitudes of the channels 
$\Sigma^{0} \Xi ^{0}$ and $\Sigma^{+} \Xi ^{-}$ are in the ratio $-1:\sqrt{2}$. It is thus 
sufficient to calculate $p\Omega^{-} \to Y \Xi ^{0}$ with $Y = \Lambda, \Sigma^{0}$.
  
The normalization and sign conventions of vertices (\ref{EL1}) and (\ref{EL3}) follow to Refs. \cite{BJDR,NAGE}, the vertex (\ref{EL2}) is defined like in Ref. \cite{JONE}, the vertex (\ref{EL4}) is simply related to that of Refs. \cite{JONE,KMFF02}. We use here dimensionless coupling constants. $u$ and $u_{\mu }$ are relativistic spinors of the spin-$1/2$ particles and $\Omega ^{-}$ with the normalizations of Appendix D. Furthermore, 
\begin{eqnarray}
q &=&p_{\Omega } - p_{\Xi } = p_{Y} - p_{N}, \nonumber \\
P &=&(p_{\Omega } + p_{N })/2 = (p_{\Xi } + p_{Y})/2.
\end{eqnarray}

\begin{figure}[!htb]
\begin{center}
\includegraphics[angle=0,width=6.0 cm]{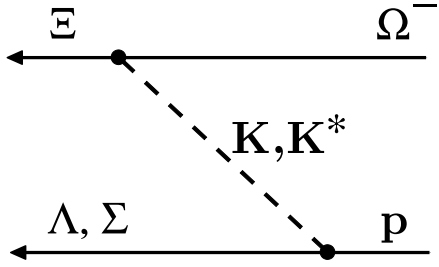}
\end{center}
\caption{Decay of $p\Omega^{-}$-exotic atom due to $t$-channel kaon exchange.}
\label{fig1}
\end{figure}

\subsection{Coupling constants}

The vertices $\Omega ^{-} \to \Xi ^{0}K^{-}$ and $\Delta ^{++} \to p\pi ^{+}$ are related by $T_{\pm }$- and $V_{\pm }$-spin operators of the $SU_{3}$
symmetry group. The coupling constants for these channels are equal in absolute values and opposite in signs. From
the width of the $\Delta \to N\pi $ decay one finds
\begin{equation}
\frac{g_{\Omega \Xi K}}{\sqrt{4\pi }} = - \frac{g_{\Delta N\pi}}{\sqrt{4\pi }} = 4.0.
\label{OXiK} 
\end{equation}
If the $SU_{3}$ relations would include the mass parameter of the vertex (\ref{EL2}), the coupling constant 
$g_{\Omega \Xi K}$ could be $m_{\Xi}/m_{N} = 1.4$ times higher. It could result to an
increase of the decay widths by a factor of two. 

Constraints from unitarity in the multichannel partial-wave analysis of the $\pi N$ 
scattering \cite{MANL} are known to be not sufficient to find relative phases of the 
amplitudes with $\pi N$ and $\pi \pi N$ final states. The relative phase of 
$g_{\Delta N\pi}$ and $g_{\Delta N\rho}$ is thus unknown. 

The ratios of $g_{N\Lambda K}$ and $g_{\Omega \Xi K}$, and $g_{N \Lambda K^{*}}$ and 
$f^{[1]}_{\Omega \Xi K^{*}}$ are known from the $SU_{6}$ symmetry (Appendix D). 

The coupling constants $g_{N \Lambda K^{*}}$ and $f_{N \Lambda K^{*}}$ are related by $SU_{3}$ to $g_{N N \rho}$ 
and $f_{N N \rho}$ the relative sign of which is fixed by the interference of the vector meson exchanges in 
the nucleon-nucleon elastic scattering, by analytical continuation of 
the $\pi N$ scattering amplitude to the $t$-channel \cite{HOHL}, and by the VMD model. 
The relative signs of $f_{\Delta N \rho}^{[i]}$ are fixed by the normalization to $\Delta$ photo- and 
electroproduction data \cite{KMFF02}. The phase of the coupling constants does affect the interference 
between the $K$ and $K^{*}$-exchange diagrams.

The vertex $\Omega ^{-} \to \Xi ^{0}K^{*-}$ is related by $SU_{3}$ with the vertex $\Delta ^{++} \to p\rho ^{+}$. 
The VMD model and isotopic symmetry allow to relate the latter with the $\Delta ^{+}\to p\gamma $
transition in which the magnetic form factor dominates.

Like for pseudoscalar mesons, the vertices $\Omega ^{-} \to \Xi ^{0}K^{*-}$ and $%
\Delta ^{++} \to p\rho ^{+}$ convert to each other by $T_{\pm }$- and $V_{\pm }$%
-spin operators. The corresponding coupling constants $f_{\Omega \Xi K^{*}}^{[i]}$ and $f_{\Delta N \rho}^{[i]}$ 
are equal in absolute values and opposite in signs. The vertex 
$\Delta^{+} \to p\rho ^{0}$ known from eVMD \cite{KMFF02} contains
isospin factor $\sqrt{2/3}$. 

In Ref. \cite{KMFF02}, the VMD model is extended in order to fulfil 
requirements of the quark counting rules by including higher radial states 
of the vector mesons. The ratios $f^{[i]}_{\Delta N V}/g_{V}$ are fixed for 
all radial states. The coupling constants $f^{[i]}_{\Delta N V}$ and $g_{V}$ 
are however known separately for the ground state $\rho$- and $\omega$-mesons 
only. In our case, the transition momenta $q^{2}$ are not high, so we apply 
an integral description of the OBE amplitude by attributing the vector meson 
exchange potential to the ground state vector mesons. The most part of the 
experimental data come from the $\Delta$ electroproduction experiments i.e. from the 
spacelike region, so behavior of the transition form factors at $q^{2} = 0$ 
is determined more reliably than at $q^{2} = m_{\rho}^{2}$. From static 
limit of the transition form factors, we obtain 
\begin{eqnarray}
\frac{f_{\Omega \Xi K^{*}}^{[1]}}{\sqrt{4\pi}} &=& - \frac{f_{\Delta N \rho}^{[1]}}{\sqrt{4\pi}} =   2.73, \label{DELTA1} \\
\frac{f_{\Omega \Xi K^{*}}^{[2]}}{\sqrt{4\pi}} &=& - \frac{f_{\Delta N \rho}^{[2]}}{\sqrt{4\pi}} = - 1.68, \label{DELTA2} \\
\frac{f_{\Omega \Xi K^{*}}^{[3]}}{\sqrt{4\pi}} &=& - \frac{f_{\Delta N \rho}^{[3]}}{\sqrt{4\pi}} = - 1.42. \label{DELTA3}
\end{eqnarray}
The pure magnetic transition would imply $f^{[1]}=-f^{[2]}=-2f^{[3]}$. As shown in Appendix D, these coupling constants 
are real and their overal sign is fixed from the requirement that the $\rho$-meson coupling constants 
$g_{NN\rho}$ and $f_{NN\rho}$ are positive. If the ground state coupling 
constants are determined as residues of the transition form factors at $q^{2} = m_{\rho}^{2}$, one could get
values 2.8 times higher. The final ambiguity should, probably, be smaller since radial states 
interfere at $q^2 =0$ destructively. In Eqs.(\ref{DELTA1}) - (\ref{DELTA3}) the $SU_{3}$ symmetry is applied to the dimensionless coupling constants.
If mass parameter of the vertex (\ref{EL4}) is included into the $SU_{3}$ scheme relations, the coupling constants 
could increase by a factor of $(m_{\Xi}/m_{N})^2 \sim 2$, while the contribution of the $K^{*}$ exchange to the decay rates
could increase by a factor of four.

The pseudoscalar coupling constants $g_{N\Lambda K}$ and $g_{N\Sigma K}$ are related by $SU_{3}$ to the pseudoscalar coupling $g_{NN\pi}$ and to the $F/D$ ratio. The $K$-exchange contribution to the 
$p\Omega ^{-} \rightarrow \Lambda \Xi^{0} , \Sigma^{0} \Xi^{0} $ decays is determined by pseudoscalar coupling constants
\begin{equation}
\frac{g_{N\Lambda K}}{\sqrt{4\pi }} = - 3.79\;\;\;\mathrm{and}\;\;\;\frac{g_{N \Sigma K}}{\sqrt{4\pi }} = 1.16. \label{ISRA1}
\end{equation}
The quoted values are from Stocks and Rijken \cite{STOC99}, model NCS97a. 
The $K^{*}$ exchange contributions to the 
$p\Omega ^{-} \rightarrow \Lambda \Xi^{0} , \Sigma^{0} \Xi^{0} $ decays are determined by vector coupling constants
\begin{eqnarray}
\frac{g_{N\Lambda K^{*}}}{\sqrt{4\pi }} &=& - 1.20\;\;\;\mathrm{and}\;\;\;\frac{g_{N \Sigma K^{*}}}{\sqrt{4\pi }} = - 0.69, \label{ISRA2} \\ 
\frac{f_{N\Lambda K^{*}}}{\sqrt{4\pi }} &=& - 3.19\;\;\;\mathrm{and}\;\;\;\frac{f_{N \Sigma K^{*}}}{\sqrt{4\pi }} = 0.32.
\label{ISRA3}
\end{eqnarray}
The model NCS97f of Ref. \cite{STOC99} gives for ${f_{N \Sigma K^{*}} }$ a value 3.5 times higher. This uncertainty does 
not affect the widths significantly, since the channel $\Sigma \Xi$ is not dominant. The uncertainties of other vector coupling constants and 
pseudoscalar coupling constants do not exceed $30\%$. 

\subsection{Transition vertices: Non-relativistic reduction}

We describe decays to the final state $\Lambda \Xi^{0}$. The decay channel 
$\Sigma^{0} \Xi^{0}$ is distinct by coupling constants and masses of the involved 
particles only. The decay amplitude for the channel $\Sigma^{+} \Xi^{-}$ 
is known from the isotopic symmetry. In what follows, we work in the rest frame of $p\Omega ^{-}$.

The energy released in the decays is small, so we apply
nonrelativistic approximation. The Coulomb interactions of the charged virtual kaons are neglected. 
The nonrelativistic reduction of vertices (\ref{EL1}) - (\ref{EL4}) gives 
\begin{eqnarray}
g_{N\Lambda K} \bar{u}(p_{\Lambda },s_{\Lambda })i\gamma _{5}u(p_{N},s_{N})
&=& \varphi _{\Lambda }^{+} C_{1}^{\alpha} q^{\alpha} \varphi_{N}, \label{EV1} \\
- \frac{g_{\Omega \Xi K}}{m_{\Xi }} \bar{u}(p_{\Xi },s_{\Xi })q_{\mu }u_{\mu
}(p_{\Omega },s_{\Omega }) &=& \varphi _{\Xi }^{+}C_{2} q^{\gamma}
\xi_{\Omega }^{\gamma},  \label{EV2} \\
\bar{u}(p_{\Lambda },s_{\Lambda})(g_{N\Lambda K^{*}}\gamma^{0 }+\frac{%
f_{N\Lambda K^{*}}}{2m_{N}}i\sigma^{0 \nu }q_{\nu })u(p_{N},s_{N})&=&
\varphi^{+}_{\Lambda} C_{3} \varphi_{N},  \label{sJ3} \\
\bar{u}(p_{\Lambda },s_{\Lambda})(g_{N\Lambda K^{*}}\gamma^{\alpha}+\frac{%
f_{N\Lambda K^{*}}}{2m_{N}}i\sigma^{\alpha \nu }q_{\nu })u(p_{N},s_{N})&& \label{sJ4} \\
= \varphi^{+}_{\Lambda}(C_{4} p_{\Xi}^{\alpha} &+& C_{5}^{\alpha \beta} q^{\beta}) \varphi_{N},  \nonumber  \\
- i \sum_{i} \frac{f_{\Omega \Xi K^{*}}^{[i]}}{m_{\Xi}^{2}} \bar{u}(p_{\Xi },s_{\Xi }) \bar{\Gamma}_{i \nu}^{0}u_{\nu
}(p_{\Omega },s_{\Omega })&& \label{sJ1} \\ 
= \varphi ^{+}_{\Xi} ( C_{6}^{\alpha \beta \gamma} q^{\alpha} p_{\Xi}^{\beta} &+& C_{7}^{\alpha \beta \gamma} q^{\alpha}
q^{\beta})\xi^{\gamma}_{\Omega},   \nonumber \\
- i\sum_{i} \frac{f_{\Omega \Xi K^{*}}^{[i]}}{m_{\Xi}^{2}} \bar{u}(p_{\Xi },s_{\Xi }) \bar{\Gamma}_{i \nu}^{\alpha}
u_{\nu}(p_{\Omega },s_{\Omega }) &=& \varphi ^{+}_{\Xi} C_{8}^{\alpha
\beta \gamma} q^{\beta} \xi^{\gamma}_{\Omega},  \label{sJ2}
\end{eqnarray}
where 
\begin{eqnarray}
C_{1}^{\alpha} &=& - \frac{ig_{N\Lambda K}}{2m_{N}} \sigma^{\alpha}, \\
C_{2} &=& \frac{g_{\Omega \Xi K}}{m_{\Xi }}, \\
C_{3} &=& g_{N\Lambda K^{*}}, \\
C_{4} &=& - \frac{g_{N\Lambda K^{*}}}{m_{N}}, \\
C_{5}^{\alpha \beta} &=& \frac{1}{2m_{N}}( - g_{N\Lambda
K^{*}}\delta^{\alpha \beta} + (g_{N\Lambda K^{*}} + f_{N\Lambda K^{*}})
i\epsilon^{\alpha \gamma \beta} \sigma^{\gamma}), \\
C_{6}^{\alpha \beta \gamma} &=& - i \frac{f_{\Omega \Xi K^{*}}^{[1]}}{m_{\Xi}^{2}} ( \sigma^{\alpha} \delta^{\beta \gamma} - \sigma^{\beta} \delta^{ \alpha \gamma}), \\
C_{7}^{\alpha \beta \gamma} &=& - i \frac{f_{\Omega \Xi K^{*}}^{[1]} + f_{\Omega \Xi K^{*}}^{[2]}}{2 m_{\Xi}^{2}}\sigma^{\alpha}
\delta^{\beta \gamma}, \\
C_{8}^{\alpha \beta \gamma} &=& - m_{\Omega} C_{6}^{\alpha \beta \gamma}.
\end{eqnarray}
Here, the terms $(m_{\Omega }-m_{\Xi })/m_{\Xi }\sim (m_{\Lambda }-m_{N})/m_{N} \sim \mathbf{q}^{2}/m_{N}^{2} \sim (v/c)^{2}$ are systematically neglected. 

In general, the $N\Lambda$ vector current is not conserved. Its longitudinal component, however, does not contribute 
to the decay amplitude, since the divergence of the $\Omega \Xi$ vector current defined by (\ref{EL4}) vanishes both 
on- and off-shell. It allows to calculate the $K^{*}$ exchange amplitude as the product of $N\Lambda$ and $\Omega \Xi$ 
vector currents. In the limit of $m_N = m_{\Lambda}$ the current (\ref{EL3}) is conserved, so its divergence $\sim (m_{\Lambda }-m_{N})/m_{N}$ can be neglected in the nonrelativistic limit anyway.

\subsection{Decays of $p\Omega ^{-}$ atoms}

The $p\Omega ^{-}$ atomic state is defined by 
\begin{equation}
|\mathbf{P,}s_{N3},s_{\Omega 3},nlm>=\int \frac{d\mathbf{p}_{N}d\mathbf{p}%
_{\Omega }}{(2\pi )^{3}}\delta (\mathbf{P}-\mathbf{p}_{N}-\mathbf{p}_{\Omega
})\Psi _{nlm}(\frac{\mathbf{p}_{\Omega }-\mathbf{p}_{N}}{2})|\mathbf{p}%
_{N},s_{N3};\mathbf{p}_{\Omega },s_{\Omega 3}>.
\nonumber
\end{equation}
The final state consists of a plane wave of $\Lambda $ and $\Xi $: 
\begin{equation}
|\mathbf{p}_{\Lambda }\mathbf{,}s_{\Lambda 3};\mathbf{p}_{\Xi },s_{\Xi 3}>.
\end{equation}

These states are normalized by 
\begin{eqnarray}
<\mathbf{P}^{\prime }\mathbf{,}s_{N3}^{\prime },s_{\Omega 3}^{\prime
},n^{\prime }l^{\prime }m^{\prime }&|&\mathbf{P,}s_{N3},s_{\Omega 3},nlm> \nonumber \\
&&=(2\pi )^{3}\delta (\mathbf{P}^{\prime }-\mathbf{P})\delta
_{s_{N3}^{\prime },s_{N3}}\delta _{s_{\Omega 3}^{\prime },s_{\Omega
3}}\delta _{n^{\prime }n}\delta _{l^{\prime }l}\delta _{m^{\prime }m}, \nonumber \\
<\mathbf{p}_{\Lambda }^{\prime }\mathbf{,}s_{\Lambda 3}^{\prime };\mathbf{p}%
_{\Xi }^{\prime },s_{\Xi 3}^{\prime }&|&\mathbf{p}_{\Lambda }\mathbf{,}%
s_{\Lambda 3};\mathbf{p}_{\Xi },s_{\Xi 3}> \nonumber \\
&&=(2\pi )^{3}\delta (\mathbf{p}_{\Lambda }^{\prime }\mathbf{-p}_{\Lambda })(2\pi )^{3}\delta (\mathbf{p}%
_{\Xi }^{\prime }-\mathbf{p}_{\Xi })\delta _{s_{\Lambda 3}^{\prime
},s_{\Lambda 3}}\delta _{s_{\Xi 3}^{\prime },s_{\Xi 3}}. \nonumber
\end{eqnarray}
The first condition is a consequence of the orthogonality of hydrogen-like
wave functions
\begin{equation}
\int \frac{d\mathbf{q}}{(2\pi )^{3}}\Psi _{n^{\prime }l^{\prime }m^{\prime
}}(\mathbf{q})^{*}\Psi _{nlm}(\mathbf{q})=\delta _{n^{\prime }n}\delta
_{l^{\prime }l}\delta _{m^{\prime }m}.
\end{equation}

Consider the lowest order $S$-matrix element responsible for decay $p\Omega
^{-}\rightarrow \Lambda \Xi ^{0}$: 
\begin{eqnarray}
S_{fi} &=&i^{2}\int \int d^{4}xd^{4}y<\mathbf{p}_{\Lambda },s_{\Lambda 3};%
\mathbf{p}_{\Xi },s_{\Xi 3}|
T[\mathcal{L}_{\Xi \Omega  K}(x)\mathcal{L}_{\Lambda N K}(y) + \nonumber \\ 
&& \mathcal{L}_{\Xi \Omega  K^{*}}(x)\mathcal{L}_{\Lambda N K^{*}}(y)]
|\mathbf{0,}s_{p3},s_{\Omega 3},nlm> = i(2\pi )^{4}\delta ^{4}(P_{f}\mathbf{-}P_{i})\frak{M}_{fi}. \nonumber
\end{eqnarray}
Entering this expression 
\begin{eqnarray}
\mathcal{L}_{\Lambda N K}(x) &=& \phi^{+}(x)J_{\Lambda N}(x)    , \\
\mathcal{L}_{\Xi \Omega K}(x) &=& J_{\Xi \Omega}(x) \phi(x), \\
\mathcal{L}_{\Lambda N K^{*}}(x) &=& - \phi^{+}_{\mu}(x) J^{\mu}_{\Lambda N}(x), \\
\mathcal{L}_{\Xi \Omega K^{*}}(x) &=& - J^{\mu}_{\Xi \Omega}(x) \phi_{\mu}(x).
\end{eqnarray} 
are effective Lagrangian densities corresponding to vertices (\ref{EL1}) - (\ref{EL4}), 
$\phi(x)$ and $\phi_{\mu}(x)$ are $K^{+}$- and $K^{+*}$-meson fields.

In the nonrelativistic approximation, the amplitude takes the form 
\begin{equation}
\frak{M}_{fi} = \sum_{A} \frak{M}_{fi}^{A}
\end{equation}
where
\begin{eqnarray}
\frak{M}_{fi}^{1} &=& - [\varphi _{\Lambda }^{+}C_{1}^{\alpha }\varphi
_{N}][\varphi _{\Xi }^{+}C_{2}J_{P}^{\alpha \gamma }\xi _{\Omega }^{\gamma }], \\
\frak{M}_{fi}^{2} &=& [\varphi _{\Lambda }^{+}C_{3}\varphi _{N}][\varphi _{\Xi
}^{+}(C_{6}^{\alpha \beta \gamma }p_{\Xi }^{\beta }J_{V}^{\alpha
}+C_{7}^{\alpha \beta \gamma }J_{V}^{\alpha \beta })\xi _{\Omega }^{\gamma }],  \\
\frak{M}_{fi}^{3} &=& - [\varphi _{\Lambda }^{+}(C_{4}p_{\Xi }^{\alpha }J_{V}^{\lambda
}+C_{5}^{\alpha \beta }J_{V}^{\lambda \beta })\varphi _{N}][\varphi _{\Xi
}^{+}C_{8}^{\alpha \lambda \gamma }\xi _{\Omega }^{\gamma }]. \label{9999}
\end{eqnarray}
The first matrix element corresponds to the $K$-meson exchange, the last two ones correspond 
to the $K^{*}$-meson exchanges as the products of timelike and spacelike components of 
the transition vector currents. 

The functions $J_{M}^{\alpha _{1}...\alpha _{s}}$ entering Eqs.(\ref{9999}) $s=1,2$ appear upon the integration over 
the atomic wave function:
\begin{equation}
J_{M}^{\alpha _{1}...\alpha _{s}}(\mathbf{p}_{\Xi }) =\int \frac{d\mathbf{p}_{N}d\mathbf{p}%
_{\Omega }}{(2\pi )^{3}}\delta (\mathbf{p}_{N}+\mathbf{p}_{\Omega })\Psi
_{nlm}(\frac{\mathbf{p}_{\Omega }-\mathbf{p}_{N}}{2})\frac{1}{q^{2}-m_{M}^{2}%
}q^{\alpha _{1}}...q^{\alpha _{s}}
\end{equation}
where ${q}={p}_{\Omega }-{p}_{\Xi }={p}_{\Lambda }-{p}_{N}$ and $M=P,V$. These functions are symmetric in $\alpha _{1}...\alpha _{s}$ and 
can be converted to the form
\begin{eqnarray*}
J_{M}^{\alpha _{1}...\alpha _{s}}(\mathbf{p}_{\Xi }) &=& \int \frac{d\mathbf{p}_{\Omega }}{(2\pi )^{3}}\Psi _{nlm}(\mathbf{p}%
_{\Omega })\frac{1}{q^{2}-m_{M}^{2}}q^{\alpha _{1}}...q^{\alpha _{s}} \\
&=&\int \int \frac{d\mathbf{q}}{(2\pi )^{3}}d\mathbf{x}\Psi _{nlm}(\mathbf{x}%
)e^{-i(\mathbf{q}+\mathbf{p}_{\Xi })\mathbf{x}}\frac{1}{q^{2}-m_{M}^{2}}%
q^{\alpha _{1}}...q^{\alpha _{s}} \\
&=&\int d\mathbf{x}\Psi _{nlm}(\mathbf{x})e^{-i\mathbf{p}_{\Xi }\mathbf{x}%
}\int \frac{d\mathbf{q}}{(2\pi )^{3}}q^{\alpha _{1}}...q^{\alpha _{s}}e^{-i%
\mathbf{qx}}\frac{1}{q^{2}-m_{M}^{2}} \\
&=&-\int d\mathbf{x}\Psi _{nlm}(\mathbf{x})e^{-i\mathbf{p}_{\Xi }\mathbf{x}}i%
\frac{\partial }{\partial x^{\alpha _{1}}}...i\frac{\partial }{\partial
x^{\alpha _{s}}}\frac{e^{-m_{M}^{*}r}}{4\pi r}.
\end{eqnarray*}
Here, use is made of the momentum conservation $\mathbf{p}_{N}+\mathbf{p}%
_{\Omega }=\mathbf{p}_{\Lambda }+\mathbf{p}_{\Xi }=\mathbf{0}$. The bound
state wave function is written then in the coordinate representation. The
effective meson mass entering the Yukawa potential equals $m_{M}^{*}=\sqrt{%
m_{M}^{2}-q_{0}^{2}}$ where 
\[
q_{0}=\left( m_{\Omega }+\frac{1}{2}E_{n}\right) -E_{\Xi }=-\left( m_{N}+%
\frac{1}{2}E_{n}\right) +E_{\Lambda }=\frac{\sqrt{s}(m_{\Omega
}-m_{N})-m_{\Xi }^{2}+m_{\Lambda }^{2}}{2\sqrt{s}} 
\]
and $\sqrt{s}=m_{\Omega }+m_{N}+E_{n}.$ The effective masses of $K$- and $K^{*}$-mesons, $m_{K}^{*} = 410$ MeV and $m_{K^{*}}^{*} = 850$ MeV, are distinct from the vacuum masses $m_{K} = 494$ MeV and $m_{K^{*}} = 892$ MeV.

We wish to find
\begin{equation}
\sum_{spins}\left| \frak{M}_{fi}\right| ^{2}=\sum_{A,B} \mathcal{R}_{AB},
\end{equation}
where
\[
\mathcal{R}_{AB} \equiv \sum_{spins} \frak{M}_{fi}^{A} \frak{M}_{fi}^{B*}.
\]

The straightforward calculations give
\begin{eqnarray}
\mathcal{R}_{11} &=&\frac{8}{3} \frac{g_{N\Lambda K}^2}{4m_{N}^2} \frac{g_{\Omega \Xi K}^{2}}{m_{\Xi }^2}J_{P}^{\alpha \lambda }J_{P}^{\alpha
\lambda *}, \label{R11} \\
\mathcal{R}_{12} &=& \mathcal{R}_{21}^{*}=0, \label{R12} \\
\mathcal{R}_{13} &=& \mathcal{R}_{31}^{*} = - \frac{8}{3}\frac{g_{N\Lambda K}}{2m_{N}}\frac{g_{\Omega \Xi K}}{%
m_{\Xi }}\frac{g_{N\Lambda K^{*}}+f_{N\Lambda K^{*}}}{2m_{N}} \frac{m_{\Omega} f_{\Omega \Xi K^{*}}^{[1]}}{m_{\Xi}^{2}} \nonumber \\
&&\;\;\;\;\;\;\;\;\;\;\;\;\;\; \times ( J_{P}^{\alpha \lambda } J_{V}^{\alpha \lambda *} - J_{P}^{\alpha \alpha} J_{V}^{\lambda \lambda *}), \label{R13} \\
\mathcal{R}_{22} &=& \frac{8}{3} \frac{g_{N\Lambda K^{*}}^2}{{m_{\Xi}^{4}}} ( f_{\Omega \Xi K^{*}}^{[1] 2}(p_{\Xi}^{2}J_{V}^{\alpha} J_{V}^{\alpha *} - p_{\Xi}^{\alpha} J_{V}^{\alpha} p_{\Xi}^{\lambda} J_{V}^{\lambda *} ) \nonumber \\
&&\;\;\;\;\;\;\;\;\;\;\;\;\;\;+ \frac{1}{8}(f_{\Omega \Xi K^{*}}^{[1]} + f_{\Omega \Xi K^{*}}^{[2]})^2 (3J_{V}^{\alpha \lambda}J_{V}^{\alpha \lambda *} - J_{V}^{\alpha \alpha}J_{V}^{\lambda \lambda *})), \label{R22} \\
\mathcal{R}_{23} &=& \mathcal{R}_{32}^{*} = \frac{16}{3}\frac{g_{N\Lambda K^{*}}^{2}}{2m_{N}} 
\frac{m_{\Omega} f_{\Omega \Xi K^{*}}^{[1] 2}}{ m_{\Xi}^{4}}(p_{\Xi }^{2} J_{V}^{\alpha }J_{V}^{\alpha *} - p_{\Xi }^{\alpha
}J_{V}^{\alpha }p_{\Xi }^{\lambda }J_{V}^{\lambda *}), \label{R23} \\
\mathcal{R}_{33} &=&\frac{2m_{\Omega }^{2}}{3m_{N}^{2}} \frac{ f_{\Omega \Xi K^{*}}^{[1] 2}}{ m_{\Xi}^{4}} ( 4g_{N\Lambda
K^{*}}^{2}(p_{\Xi }^{2} J_{V}^{\alpha }J_{V}^{\alpha *} - p_{\Xi }^{\alpha
}J_{V}^{\alpha }p_{\Xi }^{\lambda }J_{V}^{\lambda *}) \nonumber \\
&&\;\;\;\;\;\;\;\;\;\;\;\;\;\;+(g_{N\Lambda
K^{*}}+f_{N\Lambda K^{*}})^{2}(J_{V}^{\alpha \lambda }J_{V}^{\alpha \lambda *} + J_{V}^{\alpha \alpha }J_{V}^{\lambda \lambda *})) . \label{R33}
\end{eqnarray}
The values $\mathcal{R}_{11}$, $\mathcal{R}_{22}$, $\mathcal{R}_{23}$, and $\mathcal{R}_{33}$ are not affected by phase
ambiguities of the coupling constants. The phase factor of the interference 
term $\mathcal{R}_{13}$ is fixed in Appendix D. 

The functions $J_{M}^{\alpha_{1}...\alpha_{s}}$ for $s=0,1,2$ can be expressed as
follows: 
\begin{eqnarray}
J_{M}(\mathbf{p}) &=& (-i)^{l}Y_{lm}(\frac{\mathbf{p}}{p}%
)J_{M}^{[0]}(p),  \label{J0} \\
J_{M}^{\alpha }(\mathbf{p}) &=& \frac{\partial }{\partial p^{\alpha }} (-i)^{l}Y_{lm}(\frac{\mathbf{p}}{p})J_{M}^{[1]}(p),
\label{J1} \\
J_{M}^{\alpha \beta }(\mathbf{p}) &=& \left( \frac{\partial }{\partial p^{\alpha }}\frac{\partial }{\partial
p^{\beta }} - \frac{1}{3}\delta ^{\alpha \beta }\frac{\partial }{\partial
p^{\gamma }}\frac{\partial }{\partial p^{\gamma }}\right) (-i)^{l}Y_{lm}(\frac{%
\mathbf{p}}{p})J_{M}^{[2]}(p)\nonumber \\&&+\frac{1}{3}\delta ^{\alpha \beta }(-i)^{l}Y_{lm}(%
\frac{\mathbf{p}}{p})J_{M}^{[3]}(p)  \label{J2}
\end{eqnarray}
where 
\begin{eqnarray}
J_{M}^{[0]}(p) &=& - \int_{0}^{+\infty }r^{2}drR_{nl}(r)j_{l}(pr)\frac{%
e^{-m_{M}^{*}r}}{r},  \label{J00} \\
J_{M}^{[1]}(p) &=& - \int_{0}^{+\infty }r^{2}drR_{nl}(r)\left(
j_{l}(pr)-\delta _{l0}\right) \left( \frac{m_{M}^{*}}{r}+\frac{1}{r^{2}}%
\right) \frac{e^{-m_{M}^{*}r}}{r},  \label{J11} \\
J_{M}^{[2]}(p) &=& - \int_{0}^{+\infty }r^{2}drR_{nl}(r)\left( j_{l}(pr)-(1-%
\frac{1}{6}p^{2}r^{2})\delta _{l0}-\frac{1}{3}pr\delta _{l1}\right) \nonumber \\ 
&&\;\;\;\;\;\;\;\;\;\;\;\;\;\;\;\;\;\; \times \left( 
\frac{m_{M}^{*2}}{r^{2}}+\frac{3m_{M}^{*}}{r^{3}}+\frac{3}{r^{4}}\right) 
\frac{e^{-m_{M}^{*}r}}{r},  \label{J22} \\
J_{M}^{[3]}(p) &=& - m_{M}^{*2}J_{M}^{[0]}(p)-R_{nl}(0).  \label{J33}
\end{eqnarray}
It is customary to remove $\delta(\mathbf{x})$ terms originating from 
meson exchanges from baryon-baryon potentials, since such 
terms are removed anyway by zeros in the baryon wave functions in presence 
of the repulsive core. We thus drop the $R_{nl}(0)$ term in Eq.(\ref{J33}).

To arrive at Eqs.(\ref{J00}) - (\ref{J33}), we used decomposition of the plane wave 
\[
e^{-i\mathbf{px}}=4\pi \sum_{lm} i^{-l}j_{l}(pr)Y_{lm}(\frac{\mathbf{p}}{p}%
)Y_{lm}^{*}(\frac{\mathbf{x}}{r}) 
\]
and decomposition of the atomic wave function 
\[
\Psi _{nlm}(\mathbf{x}) = R_{nl}(r) Y_{lm}(\frac{\mathbf{x}}{r}). 
\]

The integrand of $J_{M}^{[1]}(p)$ is regularized to insure the possibility of interchanging 
order of the integration over $r$ and the differentiation over ${p}^{\alpha}$. 
The replacement $j_{l}(pr) \rightarrow j_{l}(pr)-\delta _{l0}$ does not 
affect the result, since for $l=0$ $J_{M}^{[1]}(p)^{\prime}$ enters Eqs.(\ref{I1}) - (\ref{I4}) 
only. The integrand of $J_{M}^{[2]}(p)$ is regularized at $r=0$ similarly. The functions $J_{M}^{[0]}(p)$ and 
$J_{M}^{[3]}(p)$ are well defined. 

The functions (\ref{J0}) - (\ref{J2}) have dimensions (in units 
$\hbar = c = 1$): 
$\lbrack J_{M}^{\alpha_{1} ... \alpha_{s}}(\mathbf{p})] = \lbrack M^{-1/2 +s} ]$, 
and also 
$\lbrack J_{M}^{[0]}(p)] = \lbrack M^{-1/2}]$, 
$\lbrack J_{M}^{[1]}(p)] = \lbrack M^{3/2}]$, 
$\lbrack J_{M}^{[2]}(p)] = \lbrack M^{7/2}]$, and 
$\lbrack J_{M}^{[3]}(p)] = \lbrack M^{3/2}]$.

By integrating the decay amplitude squared over ${p}_{\Xi}$, four integrals appear
\begin{eqnarray}
K_{MM^{\prime}}^{[1]}(p) &=& \int d\Omega _{\mathbf{p}}J_{M}^{\alpha \lambda}(\mathbf{p})J_{M^{\prime}}^{\alpha \lambda *}(\mathbf{p}),  \label{I1} \\
K_{MM^{\prime}}^{[2]}(p) &=& \int d\Omega _{\mathbf{p}}J_{M}^{\alpha \alpha}(\mathbf{p})J_{M^{\prime}}^{\lambda \lambda *}(\mathbf{p}),  \label{I2} \\
K_{MM^{\prime}}^{[3]}(p) &=& \int d\Omega _{\mathbf{p}}p^{\alpha} J_{M}^{\alpha}(\mathbf{p}) p^{\lambda} J_{M^{\prime}}^{\lambda *}(\mathbf{p}), \label{I3} \\
K_{MM^{\prime}}^{[4]}(p) &=& p^{2}\int d\Omega _{\mathbf{p}}J_{M}^{\alpha}(\mathbf{p}) J_{M^{\prime}}^{\alpha *}(\mathbf{p}). \label{I4}
\end{eqnarray}
In Appendix E, we provide identities useful for the integration over the angles.
The integrals (\ref{I1}) - (\ref{I4}) are found to be 
\begin{eqnarray*}
K_{MM^{\prime}}^{[1]}(p) &=& J_{M}^{[2]}(p)^{\prime \prime }J_{M^{\prime}}^{[2]}(p)^{\prime \prime} 
+ 2\frac{l^{2}+l+1}{p^{2}}J_{M}^{[2]}(p)^{\prime }J_{M^{\prime}}^{[2]}(p)^{\prime} -  3\frac{l(l+1)}{p^{3}}J_{M}^{[2]}(p)J_{M^{\prime}}^{[2]}(p)^{\prime }\\
&-& 3\frac{l(l+1)}{p^{3}}J_{M}^{[2]}(p)^{\prime } J_{M^{\prime}}^{[2]}(p)
 + \frac{l(l+1)(l^{2}+l+1)}{p^{4}}J_{M}^{[2]}(p)J_{M^{\prime}}^{[2]}(p)\\&&-\frac{1}{3}%
\Delta _{l}J_{M}^{[2]}(p)\Delta _{l}J_{M^{\prime}}^{[2]}(p)+\frac{1}{3}%
J_{M}^{[3]}(p)J_{M^{\prime}}^{[3]}(p), \\
K_{MM^{\prime}}^{[2]}(p) &=&J_{M}^{[3]}(p) J_{M^{\prime}}^{[3]}(p), \\
K_{MM^{\prime}}^{[3]}(p) &=&p^{2}J_{M}^{[1]}(p)^{\prime } J_{M^{\prime}}^{[1]}(p)^{\prime }, \\
K_{MM^{\prime}}^{[4]}(p) &=&p^{2}J_{M}^{[1]}(p)^{\prime }J_{M^{\prime}}^{[1]}(p)^{\prime
}+l(l+1)J_{M}^{[1]}(p)J_{M^{\prime}}^{[1]}(p),
\end{eqnarray*}
where 
\[
\Delta_{l} = \frac{1}{p} \frac{\partial^2}{\partial p^2}p - \frac{l(l+1)}{p^2}.
\]


\begin{table}[tbp]
\centering
\addtolength{\tabcolsep}{-3pt}
\renewcommand{\arraystretch}{1.0}
\caption{
Decay widths of $p\Omega^{-}$ atoms [in keV] in the lowest atomic levels $n=1 \div 4$ and $L=0 \div 3$. 
The results are given for the $\Lambda \Xi$ and $\Sigma \Xi$ decay channels. The last channel includes 
the summation over the $\Sigma^0 \Xi^0$  and $\Sigma^+ \Xi^-$ states.
}
\vspace{2mm}
\label{tab:table3}
\begin{tabular}{|cc|c|c|c|c|}
\hline
\multicolumn{2}{|c|}{$L$} & {$0$}                   & {$1$}                   & {$2$}                    & {$3$}                     \\ \hline 
$n$ & channel             & $\Gamma \times 10^{1}$ & $\Gamma \times 10^{6}$ & $\Gamma \times 10^{12}$ & $\Gamma \times 10^{18}$      \\ \hline\hline
$1$ & $\Lambda \Xi$       & $60$                    &                         &                          &                           \\ 
$\;$& $\Sigma  \Xi$       & $1.8$                   &                         &                          &                           \\ 
$2$ & $\Lambda \Xi$       & $7.5$                   & $4.1$                   &                          &                           \\ 
$\;$& $\Sigma  \Xi$       & $0.0$                   & $2.2$                   &                          &                           \\ 
$3$ & $\Lambda \Xi$       & $2.3$                   & $1.4$                   & $2.0$                    &                           \\ 
$\;$& $\Sigma  \Xi$       & $0.0$                   & $0.8$                   & $1.7$                    &                           \\ 
$4$ & $\Lambda \Xi$       & $1.0$                   & $0.6$                   & $1.2$                    & $1.0$                     \\ 
$\;$& $\Sigma  \Xi$       & $0.0$                   & $0.4$                   & $1.0$                    & $0.6$                     \\ \hline 
\end{tabular}
\end{table}


The atomic decay width averaged over the spin states of the proton and $\Omega^-$ equals 
\begin{equation}
\Gamma = \frac{m_{\Xi} m_{\Lambda} p_{\Xi}}{32 \pi^2 \sqrt{s}} \int d\Omega_{\mathbf{p}_{\Xi}} \sum_{A,B} \mathcal{R}_{AB}, 
\label{WIDTH}
\end{equation}
where 
\begin{equation}
p_{\Xi }=\frac{\sqrt{(s-(m_{\Lambda }+m_{\Xi })^{2})(s-(m_{\Lambda }-m_{\Xi
})^{2})}}{2\sqrt{s}}
\end{equation}
is the center-of-mass $\Xi $-hyperon momentum.
In Table \ref{tab:table3} we give decay widths of lowest $p\Omega ^{-}$ atomic
levels. 

The total decay width to the channels 
$\Sigma \Xi$ is enhanced by isotopic factor 3. In $P$ states, the interference 
of the $K$ and $K^{*}$ exchnage amplitudes is destructive for the $\Lambda \Xi$ channel and constructive for the $\Sigma \Xi$ channels, 
producing thereby an additional enhancement of the $\Sigma \Xi$ channels. For $2P$ state, 
we get $\Gamma_{\Lambda \Xi} = (6.1 + 2.3 - 4.4) \times 10^{-6}$ keV and $\Gamma_{\Sigma \Xi} = (1.8 + 0.2 + 0.3) \times 10^{-6}$ keV.
The first, second and third numbers correspond to $K$, $K^{*}$ and their interference. The interference pattern of the $S$-wave is 
opposite. It is discussed in Sect. 4.5 and Appendix D. 

Since $\Omega^-$ spin is high, one can expect nontrivial $F$-dependence of the 
decay widths. The $S$-wave decay from an $F=2$ spin state e.g. can 
be suppressed, since it is coupled to a $D$-wave channel only.

In Eqs.(\ref{J00}) - (\ref{J33}) one can expand $R_{nL}(r)$ around $r=0$ and 
keep the lowest order term due to the small binding energy. As a result, the widths 
scale with $n$ approximately as squares of the $L$-th order derivatives of $R_{nL}(r)$ 
at $r=0$.

The decay width of the $2P$ state is one order of
magnitude lower as compared to $p\bar{p}$ atoms \cite{KLEM02}. The $p\bar{p}$
atoms are coupled to the continuum through the annihilation channel only. $%
p\Xi^{-}$ and $p\Sigma^{-}$ atomic states decay due to $t$-channel kaon and
pion exchanges \cite{DOGA,LOIS01}. Such decays can be calculated like decays
of $p\Omega ^{-}$ atoms.

The formalism presented here is similar to that used in Ref. \cite{SCHE} for
calculation of weak decays of a loosely bound hypothetical $H$ particle \cite
{Jaffe:1976yi}, which proceed due to kaon exchange between $\Lambda $%
-hyperons.

\subsection{Decays of high-$Z$ nucleus-$\Omega ^{-}$ atoms}

Strong decays of $\Omega ^{-}$ exotic atoms with high-$Z$ nuclei 
proceed under kinematics conditions which are more complicated as compared to $p\Omega^{-}$ decays. A microscopic approach to calculate the strong decays of hyperon $\Sigma^{-}$ atoms is discussed by Loiseau and Wycech \cite{LOIS01}. It is based on the impulse approximation where it is assumed that in the final state one has a nucleus left with a hole in a single particle state and two particles as plane waves.

The optical potential method represents a conventional approach to calculate strong decay widths 
of high-$Z$ exotic atoms. It consists in the determination of free scattering lengths and 
finding the average value of imaginary part of the nuclear optical potential. Such a method 
is in the qualitative agreement with the decay rates of $\Sigma^{-}$ exotic atoms \cite{BATT81}. 

The imaginary part of the $N\Omega^{-}$ scattering lengths 
\begin{equation}
a_{I} = \lim_{v \to 0} \frac{m_{\Omega}m_{N}} {4\pi(m_{\Omega} + m_{N})}v \sigma_{p\Omega^{-}}
\end{equation}
can be estimated using the plane wave Born approximation (PWBA) from the OBE diagram of Fig. 2:
\begin{equation}
a_{I} = 2.2 \;\mathrm{fm}
\label{SL}
\end{equation} 
both for proton and neutron. This value has the content 
$a_{I}$ = 0.45 + 1.31 + 0.09 + 0.03 + 0.44.  The first and third numbers come from the $K$-exchange, 
the second and fourth numbers come from the $K^{*}$-exchange, 
the first two numbers and the next two ones correspond to $\Lambda \Xi$ and $\Sigma \Xi$
channels, and the last number is the interference.

As proposed by M. Ericson and T. E. O. Ericson \cite{ERER66}, the atomic decay widths can be calculated 
in terms of the scattering length $a_{I}$ and integral over the nuclear density
\begin{equation}
\Gamma = \frac{4\pi}{\mu} a_{I} \int_0^R{r^2dr \rho(r) R_{nL}^2(r)}, 
\label{APPR}
\end{equation}
where $R = \sqrt{ \frac{5}{3}}<r^2>^{1/2} = r_{0}A^{1/3}$ is the nucleus radius. For $12 < A < 40$, $r_{0} = 1.35$ fm and for heavy nuclei $r_{0} = 1.20$ fm 
\cite{DOGA}. The density $\rho(r)$ is normalized according to $ \int \rho(r) dV = A$. The value of $\mu$ is the reduced mass of $\Omega^{-}$ and nucleus.

The estimate (\ref{APPR}) neglects modification of the wave function due to finite nuclear volume, relativistic effects, and real part of 
the strong interaction potential. In order to remove some of systematic errors caused by neglecting those effects, we normalize the decay width 
to $\Gamma = 0.05$ keV of $^{14}$N$\Xi^{-}$ exotic atom in $3D$ state calculated by Batty, Friedman and Gal \cite{BATT99}. 
Using Eq.(\ref{APPR}), we reproduce then with reasonable precision the reported theoretical values of the decay widths of $\Xi^{-}$-atoms formed with $^{12}$C, $^{16}$O, and $^{19}$F in $3D$ states and $^{28}$Si in $4F$ state.

The pseudoscalar coupling constant ${g_{\Xi \Lambda K}}/{\sqrt{4\pi}} = 1.10$ \cite{STOC99}, model NCS97a, is small as compared to ${g_{\Omega \Xi K}}/{\sqrt{4\pi}} = - 4.0$ given by Eq.(\ref{OXiK}). 
The channel $\Sigma \Lambda$ where the pseudoscalar coupling constant ${g_{\Xi \Sigma K}}/{\sqrt{4\pi}}= -4.69$ is
large is blocked by the energy conservation. One can expect that the $K$ exchange contribution to the $p\Omega^{-}$ scatering length is 
higher as compared to that of the $N\Xi^{-}$ scattering length. 
The amplitudes $p\Omega^{-} \to \Lambda \Xi$ due to the $K$ and $K^{*}$ exchanges are in the ratio 
$1:1.7$. 

The imaginary part of the $N\Xi^{-} \to \Lambda \Lambda$ scattering length is only $a_{I} = 0.04$ fm \cite{BATT99}. 
This small value is the result of small coupling constant ${g_{\Xi \Lambda K}}$, statistical weight 1/16 of the $\Lambda \Lambda$ channel, 
and small phase space of $\Lambda \Lambda$. Increasing the imaginary potential causes the calculated widths to increase by roughly the same proportion. Taking into account the decreased Bohr radius due to the higher mass of $\Omega^{-}$ and significantly increased imaginary part of the scattering length, we obtain decay widths of the $\Omega^{-}$ exotic atoms reported in Table 
\ref{tab:table5}. Using the same assumptions, we obtain for $^{208}$Pb$\Omega^{-}$ atom 
$\Gamma_9 \sim 7$ keV in $L =n-1=9$ state, 
$\Gamma_{10} \sim 0.1$ keV in $L =n-1=10$ state, and 
$\Gamma_{11} \sim 2$ eV for $L =n-1=11$. 
In Table \ref{tab:table5}, we restricted by giving estimates for nuclei $A \ge 6$, 
since Eq.(\ref{APPR}) matches smoothly at $A=1$ with decay widths of $S$ levels, whereas we list decay widths of $P$ levels.

The $G$-matrix formalism by Yamamoto \textit{et al.} \cite{YAMA94} results to decay 
widths of $\Xi$ atoms one order of the magnitude higher \cite{BATT99}. Accordingly, one
can consider above estimates of widths as the lower bounds. 

The value of the scattering length Eq.(\ref{SL}) is three times higher as compared to the $pK^-$ scattering length. 
Uncertainties in the coupling constants of $\Omega^{-}$, discussed in Sect. 4.2, point in many cases towards even higher values of $a_I$. 

We see from Table \ref{tab:table5} that widths of the $2P$ states are several orders of the magnitude 
higher than those required to differentiate the contact $P$-wave interactions and up to three orders of magnitude 
higher than the long-range interactions. The kaon exchange in circular orbits of the $^{208}$Pb$\Omega^{-}$ exotic 
atoms has negligible effect starting from $L \sim 10$. 

The strong-interaction shift is usually expected to be as large as the width and a fraction of the strong-interaction shift could also be spin-dependent, thus contributing to the apparent hyperfine splitting. Lorentz scalar component of the two-kaon 
exchange potential does not generate hyperfine splitting. The vector component generates hyperfine splitting due to spin-tensor interaction $\Delta E \sim (\alpha Z)^{2} \times 1$ MeV $\sim 3$ keV (for mass and spin dependence of OBE potentials of spin-1/2 particles see Ref. \cite{GROM}, the numerical values are quoted for $^{14}N \Omega^{-}$). The effect is comparable with the long-range quadrupole-orbit interaction.

In Ref. \cite{GIAN}, the circular transitions in $^{208}$Pb$\Omega^{-}$ exotic atoms for $L=10 \to 9$ levels are estimated. The transition energy is about 0.5 MeV, each level splits into four sublevels due to spin-orbit ($\sim 2$ keV for $L=9$) and quadrupole-orbit ($\sim 0.2$ keV for $L=9$ and $Q \sim 3 \times 10^{-2}$ fm$^2$) interactions. In the experiments with $\Sigma^{-}$ atoms \cite{HERT}, the peak positions of the photon energies are determined with accuracy of a few tens of eV. Using the same technique, it would be possible to measure the $\Omega^{-}$ quadrupole moment with an accuracy of $\sim 30 \%$ in the circular transitions $L=11 \to 10$. The strong interaction shift of the lower 
$L=10$ level is expected to be $\sim \Gamma_{10}$, while splitting due to strong interactions, sensitive to the quadrupole moment, is smaller: 
$\Delta E \sim (\alpha Z/n)^{2} \Gamma_{10} < 1$ eV.

In Appendix C, we compare the natural widths and the contact $P$-wave interactions in $\mu$-meson exotic 
atoms. The estimates reported in Table \ref{tab:table91} 
suggest that in high-$Z$ nuclei the contact interactions are of order of the radiative widths of 
$2P_{3/2}$ levels.

\section{Conclusions}
\setcounter{equation}{0}

In this work, we investigated the possibility of measurement of the $\Omega^{-}$
quadrupole moment by observing X-rays from low-$L$ transitions in $\Omega^{-}$ exotic atoms. The magnitude 
of fine and hyperfine splittings of the energy levels has been compared to strong decay widths caused by
reactions $N\Omega^{-} \to \Lambda \Xi, \Sigma \Xi$ in $\Omega^{-}$ atoms formed with light stable nuclei 
with atomic numbers below 10 and spins above $1/2$.

We proposed, firstly, a minor modification of the $\Omega^{-}$ spin-orbit interaction used 
in the earlier works \cite{GOLD,KARL2} in order to bring it in agreement with 
theory of Thomas precession. 

Secondly, we described new kinds of the contact $P$-wave interactions for particles with 
electric quadrupole and magnetic dipole moments. We found that Thomas correction for
the quadrupole-spin contact $P$-wave interaction is the same as for the spin-orbit 
long-range interaction. The Darwin term connected to the particles Zitterbewegung 
represents yet another source of contact $P$-wave interactions.

The long-range interactions appearing to the order $(v/c)^2$ such as spin-orbit interactions, 
spin-spin tensor interaction, and quadrupole - orbit interactions, have been discussed and 
included into the numerical estimates of the energy splitting. We showed the for $2P$ states 
of $\Omega^{-}$ exotic atoms with stable nuclei up to $^{19}$F the contact interactions are $2-5$ 
orders of magnitude weaker than the conventional long-range interactions.
The quantitative evaluation of the contact $P$-wave interactions suffers from the poor 
knowledge of the short-range component of the baryon-baryon interactions.

The contact $P$-wave electric quadrupole - magnetic dipole interaction exists in ordinary atoms 
and $\mu$-meson atoms. In high-$Z$ nuclei, the magnitudes of the contact $P$-wave interactions 
in $\mu$-meson atoms are comparable with the natural widths of $2P_{3/2}$ levels.

Thirdly, we calculated strong decay rates of $p\Omega^{-}$ exotic atoms due to reactions 
$p\Omega^{-} \to \Lambda \Xi, \Sigma \Xi$ caused by $K$ and $K^*$ $t$-channel 
exchanges. The problem is solved analytically for arbitrary principal and orbital 
quantum numbers. The decay rates of the lowest $p\Omega^{-}$ atomic levels, 
averaged over the proton and $\Omega^{-}$ spin states, are reported 
in Table \ref{tab:table3}.

Rough estimates of strong decay rates of the $\Omega^{-}$ exotic atoms formed with the light nuclei and of the
$^{208}$Pb$\Omega^{-}$ exotic atom have been made. For $2P$ states of the low-$Z$ $\Omega^{-}$ atoms, 
we get strong decay widths up to three orders of magnitude higher than splitting caused by the conventional 
long-range interactions and, respectively, 5-6 orders of magnitude higher than the contact $P$-wave 
interactions. Table \ref{tab:table5} summarizes the estimates of magnitudes of the 
interactions and the strong decay widths. The contact $P$-wave interactions are 
not useful for measurement of the $\Omega^{-}$ quadrupole moment. 

Strong decay channels in $^{208}$Pb$\Omega^{-}$ exotic atoms are small in the circular transitions 
starting from $L = n-1 \sim 10$. Such transitions minimize theoretical uncertainties inherent to 
the problem and can be suitable for measurements of the $\Omega^{-}$ quadrupole moment.

The authors are grateful to M. D. Semon for correspondence and B. V. Martemyanov for discussions on 
the Thomas precession effect. The authors wish to acknowledge referees of Nuclear Physics A 
for useful remarks and suggestions. This work is supported by RFBR grant No. 06-02-04004 and DFG grant 
No. 436 RUS 113/721/0-2.

\begin{appendix}

\section{Thomas precession}
\setcounter{equation}{0}

A formally complete treatment of the contact $P$-wave interactions
would require the knowledge of other interactions and, specifically, 
the rate of Thomas precession to order $(v/c)^4$. There exists a controversy 
in the evaluation of Thomas precession effect beyond $(v/c)^2$ as discussed 
recently \cite{MALY06}. Here, we give relativistic treatment of Thomas
precession.

Let us consider coordinate systems $K$ and $K^{\prime }$ in which a particle 
has four-velocities $u=(\gamma,\gamma \mathbf{v})$ and $u^{\prime}=(1, \mathbf{0})$, 
respectively. In the coordinate system $K$, particle moves with velocity 
$\mathbf{v}$, whereas in the coordinate system $K^{\prime }$ it
is at rest. Given that $w=(w_{0},\mathbf{w})$ in $K$, we search for the coordinates of $w^{\prime}$ in
$K^{\prime }$.

We split $\mathbf{w}$ into the parallel and transverse components with respect to the direction of the velocity: 
$\mathbf{w}_{||}=\mathbf{n(nw)}$ and $\mathbf{w}_{\perp }=\mathbf{w-n(nw),}$ 
where $\mathbf{\;n=v}/v.$ The transformation of $(w^{0},w_{||})$ where 
$w_{||}=\mathbf{(nw)}$, are well known while $\mathbf{w}_{\perp }$ does not transform. One
can write therefore 
\begin{eqnarray}
w_{0}^{\mathbf{\prime }} &=&\gamma (w_{0}-v\mathbf{nw}), \\
\mathbf{nw}^{\mathbf{\prime }} &=&\gamma (\mathbf{nw}-vw_{0}), \\
\mathbf{w}^{\prime }-\mathbf{n(nw}^{\prime }) &=&\mathbf{w}-\mathbf{n(nw}).
\end{eqnarray}
This system of equations allows to find the Lorentz transformation matrix (see e.g. \cite{WIKILT})
\begin{equation}
L(\mathbf{v})=\left(
\begin{array}{ll}
\gamma & \mathbf{v}\gamma \\ 
\mathbf{v}\gamma & 1+(\gamma -1)\mathbf{n\otimes n}
\end{array}
\right).  \label{LORE}
\end{equation}
The coordinates of a four-vector $w$ in two Lorentz coordinate systems $K$ and $K^{\prime}$
are related by 
\begin{equation}
L(\mathbf{v})w^{\prime}=w.
\label{LOTR}
\end{equation}
Remind that $K^{\prime}$ moves with velocity $\mathbf{v}$ in $K$.

The particle polarization is a three-dimensional unit vector, $\mathbf{a}$. It is defined in the particle rest frame. Relativistically, polarization is characterized by a four-dimensional vector $a$. Given in $K$ a particle with a four-velocity $u$ and a polarization four-vector $a$, such that
\begin{eqnarray}
u^2 &=& 1,      \label{TR1} \\
a^2 &=& -1,     \label{TR2} \\
a \cdot u &=& 0 \label{TR3},
\end{eqnarray}
one can define a three-dimensional unit vector, $\mathbf{a}$, as space-like component of $a^{\prime}$ from equation
\begin{equation}
L(\mathbf{v})a^{\prime} = a,
\label{0000}
\end{equation}
where $\mathbf{v}$ is the particle velocity in $K$. We thus make boost to $K^{\prime}$ where $u^{\prime} = (1,\mathbf{0})$, $a^{\prime} = (0,\mathbf{a})$. Three-dimensional vector $\mathbf{a}$ is called polarization of a particle which moves with velocity $\mathbf{v}$ in $K$.
 
\begin{figure}[!htb]
\begin{center}
\includegraphics[angle=0,width=4.0 cm]{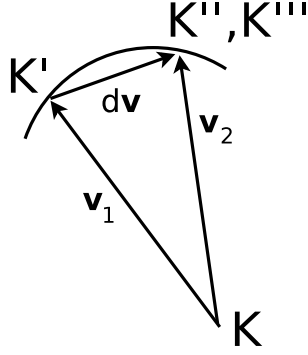}
\end{center}
\caption{Schematic representation of Lorentz boosts relating coordinate systems $K$, $K^{\prime}$, $K^{\prime \prime}$, 
and $K^{\prime \prime \prime}$ involved into calculation of the precession rate of polarization of a particle moving along a trajectory.}
\label{fig8}
\end{figure}

Let us consider particles $1$ and $2$ with four-velocities $u_1$ and $u_2$ and polarization four-vector $a_1$ and $a_2$ in $K$. They can be considered as two different states of the same particle, separated by a time interval $\delta t$ in $K$.
Particles $1$ and $2$ are at rest in coordinate systems $K^{\prime}$ and $K^{\prime \prime}$, respectively, 
as shown on Fig. \ref{fig8}. As discussed above, we can define two three-dimensional unit polarization vectors $\mathbf{a}_1$ and $\mathbf{a}_2$:
\begin{eqnarray} 
a_{1}^{\prime} &=& (0,\mathbf{a}_1) = L(-\mathbf{v}_1)a_1, \label{POLA1} \\
a_{2}^{\prime \prime} &=& (0,\mathbf{a}_2) = L(-\mathbf{v}_2)a_2, \label{POLA2}
\end{eqnarray}
velocities $\mathbf{v}_1$ and $\mathbf{v}_2$ are known since $u_1$ and $u_2$ are known.

Four-velocities $u_2$ and $u_1$ are related by a Lorentz boost. We denote $K^{\prime \prime \prime}$ a coordinate systems obtained from $K^{\prime}$ by such Lorentz boost:
$L(\delta \mathbf{v})w^{\prime \prime \prime} = w^{\prime}$. 
In particular,
\begin{eqnarray} 
L(\delta \mathbf{v})u^{\prime \prime \prime}_1 = u^{\prime}_1, \\
L(\delta \mathbf{v})u^{\prime \prime \prime}_2 = u^{\prime}_2.
\end{eqnarray}
Particle 2 is at rest both in $K^{\prime \prime}$ and $K^{\prime \prime \prime}$, $K^{\prime \prime}$ and $K^{\prime \prime \prime}$ are related by a rotation. Taking into account that $u^{\prime \prime \prime}_2 = u^{\prime}_1 = (1,\mathbf{0})$, we get
\begin{equation}
L(\delta \mathbf{v})u^{\prime}_1 = u^{\prime}_2.
\end{equation}
We require that polarization four-vectors be related by the same transformation: 
\begin{equation}
L(\delta \mathbf{v})a^{\prime}_1 = a^{\prime}_2.
\end{equation}
This ensures to fulfil Eqs.(\ref{TR1}) - (\ref{TR3}) for particle 2 provided Eqs.(\ref{TR1}) - (\ref{TR3}) are fulfilled for particle 1. The way the four-vectors are related defines \textit{parallel transport} from $K^{\prime}$ to $K^{\prime \prime \prime}$: The coordinates of all four-vectors attributed in $K^{\prime}$ and $K^{\prime \prime \prime}$, respectively, to particles $1$ and $2$ remain unchanged. In particular, $a^{\prime \prime \prime}_2 = a^{\prime}_1 = (0,\mathbf{a}_1)$.

The relativistic composition of velocities can be used to express $\mathbf{v}_2$ in terms of $\mathbf{v}_1$ and $\delta \mathbf{v}$:
\begin{equation}
\mathbf{v}_2  = \mathbf{v}_1 \oplus \delta \mathbf{v} = \mathbf{v}_1 + (\frac{1}{\gamma} - \frac{1}{\gamma + 1}\mathbf{v}_1 \otimes \mathbf{v}_1)\delta \mathbf{v}, \label{COMP1}
\end{equation}
where $\gamma = 1/\sqrt{1 - \mathbf{v}_1^{2}}$. According to an observer in $K^{\prime \prime \prime}$, $K$ moves with velocity $\mathbf{v}_3 \neq - \mathbf{v}_2$:
\begin{equation}
-\mathbf{v}_3 = \delta \mathbf{v} \oplus \mathbf{v}_1 = \mathbf{v}_1 + (1 - \mathbf{v}_1 \otimes \mathbf{v}_1)\delta \mathbf{v}. \label{COMP2}
\end{equation}
The composition of velocities is defined by the composition of Lorentz boosts: 
$u_2 = L(\mathbf{v}_1) L(\delta \mathbf{v})u^{\prime \prime \prime}_2 = L(\mathbf{v}_1 \oplus \delta \mathbf{v})u^{\prime \prime \prime}_2$. One finds that $\mathbf{v}_2 = \mathbf{v}_1 \oplus \delta \mathbf{v}$ is velocity of $K^{\prime \prime \prime}$ in $K$. $K^{\prime \prime}$ and $K^{\prime \prime \prime}$ are distinct by a rotation, so $\mathbf{v}_2$ is velocity of $K^{\prime \prime}$ in $K$ too. 

Now, it is straightforward to find
\begin{eqnarray}
a^{\prime \prime}_2 &=& L(- \mathbf{v}_2) a_2 \nonumber \\
                    &=& L(- \mathbf{v}_2) L( \mathbf{v}_1) a^{\prime}_2 \nonumber \\
                    &=& L(- \mathbf{v}_2) L( \mathbf{v}_1) L(\delta \mathbf{v}) a^{\prime \prime \prime}_2 \nonumber \\
                    &=& L(- \mathbf{v}_2) L( \mathbf{v}_1) L(\delta \mathbf{v}) a^{\prime}_1.
\end{eqnarray}
These equations show that 
\begin{equation}
\mathbf{a}_2 = \mathbb{R}\mathbf{a}_1,
\end{equation}
where $\mathbb{R}$ is a rotation matrix, such that
\begin{equation}
\left(
\begin{array}{ll}
1 & 0 \\ 
0 & \mathbb{R}
\end{array}
\right) = L(- \mathbf{v}_2) L( \mathbf{v}_1) L(\delta \mathbf{v}).
\end{equation}

Applying $\mathbb{R}$, one gets
\begin{equation}
\mathbb{R}\mathbf{a}_1 = \mathbf{a}_1 + \delta \mathbf{a}_1,
\end{equation}
where 
\begin{equation}
\delta \mathbf{a}_1 = - \frac{\gamma }{\gamma +1} ( 
\mathbf{\mathbf{v}_{1} \times \delta v}) \times \mathbf{a}_1.
\end{equation}

Circular motion where $\mathbf{v}_{1} \delta \mathbf{v} = 0$ implies 
\begin{equation}
\frac{\delta \mathbf{v}}{\gamma} = \mbox{\boldmath{$\omega$}} \times \mathbf{v}_{1}%
\delta t,
\end{equation}
where $\mbox{\boldmath{$\omega$}}$ is the orbital rotation frequency, $\delta t$ is a time interval in $K$, and $\delta\mathbf{v}/\gamma = \mathbf{v}_{2} - \mathbf{v}_{1}$ according to Eq.(\ref{COMP1}), so one gets
\begin{equation}
\delta \mathbf{a}_{1}= (1 - \gamma) \mbox{\boldmath{$\omega$}} \times \mathbf{a}_{1} \delta t.
\label{PREC}
\end{equation}
This equation shows that vector $\mathbf{a}_{1}$ experiences a precession in $K$
with frequency $\mathbf{\Omega}_{T}$ given by Eq.(\ref{thomas}). 

In Refs. \cite{RITU61,CHAK,MALY06,RITU07}, $\mathbf{\Omega}_{T}$ is $\gamma$ times smaller. The possible reason of the discrepancy
might be the noncommutativity of relativistic composition of velocities. The velocity of $K^{\prime \prime \prime}$ in $K$ and minus velocity of $K$ in $K^{\prime \prime \prime}$ deviate from $\mathbf{v}_{1}$ for circular motion by $\delta\mathbf{v}/\gamma$ and $\delta\mathbf{v}$, respectively.
The quantity $\mbox{\boldmath{$\omega$}} \times \mathbf{v}_1 \delta t$ refers to the variation of velocity of $K^{\prime \prime \prime}$ in $K$.

The definition of polarization of a moving particle using Eqs.(\ref{POLA1}) and (\ref{POLA2}) allows to attribute the transparent physical meaning to Lorentz boosts relating the coordinate systems $K \ldots K^{\prime \prime \prime}$. 
In all coordinate systems related by Lorentz boosts with the particle rest frame we observe the same three-dimensional unit polarization vector. If, however, a particle has been accelerated by a sequence of non-collinear Lorentz boosts, its polarization does rotate. The particle polarization and, accordingly, its time evolution depend on the coordinate system. From the point of view of an observer in $K^{\prime}$ there is no rotation in transit from $K^{\prime}$ to $K^{\prime \prime \prime}$. However, in $K$ we do observe a rotation.

Relativistic expression for Thomas precession frequency an external electromagnetic field
can be found using Eq.(A.22) with $\mathbf{v}_1 \times \delta\mathbf{v} = \gamma \mathbf{v}_1 \times (\mathbf{v}_{2} - \mathbf{v}_{1}) = \gamma \mathbf{v}_1 \times \dot{\mathbf{v}}_{1} \delta t$ and the covariant equation of motion for charged particles: 
\begin{equation}
\mathbf{\Omega}_{T} = - \frac{\gamma}{\gamma + 1} \frac{e}{m} \mathbf{v}_{1} \times (\mathbf{E} + \mathbf{v}_{1} \times \mathbf{B}).
\end{equation}
All quantities entering this equation are defined in the laboratory frame $K$. The Larmor precession frequency in the co-moving frame $K^{\prime}$ can be found using the Lorentz transformation for electromagnetic field: 
\begin{equation}
\mathbf{\Omega}_{L}^{\prime} = - \frac{\mu}{S}\mathbf{B}^{\prime} = - \frac{eg}{2m} (\gamma(\mathbf{B} - \mathbf{v}_{1} \times \mathbf{E})
- (\gamma - 1) \mathbf{n}_{1}(\mathbf{n}_{1} \cdot \mathbf{B})).
\end{equation}
Here, $\mathbf{n}_{1} = \mathbf{v}_{1}/|\mathbf{v}_{1}|$ and $\mathbf{B}^{\prime}$ is the magnetic field in $K^{\prime}$ (cf. Eq.(III.8)). The sum $\mathbf{\Omega}_{T} + \mathbf{\Omega}_{L}^{\prime}/\gamma$ gives the total spin precession frequency in the laboratory frame, in the exact agreement with the Bargmann-Michel-Telegdi equation \cite{LALI}. The above arguments do not rely on the assumption of $S=1/2$.

Equation (\ref{PREC}) is in agreement with Refs. \cite{MOLL52,KOBZ92,SEMO04}. It is physically equivalent 
to the equation for rotation of axes of a "Born-rigid electron" on circular orbit, derived first by F\"oppl and Daniell \cite{FOPP13}.

\section{Matrix elements of angular momentum operators and fine and hyperfine splitting in $^{14}$N$\Omega ^{-}$ exotic atom}
\setcounter{equation}{0}

Let us consider a symmetric tensor $\tau^{\alpha \beta}(\mathbf{a},\mathbf{b})$ constructed in terms of operators $\mathbf{a}$ and $\mathbf{b}$:
\begin{equation}
\tau^{\alpha \beta}(\mathbf{a},\mathbf{b}) = a^{\alpha} b^{\beta} + a^{\beta} b^{\alpha} 
- \frac{2}{3} \mathbf{a \cdot b} \delta^{\alpha \beta}. \label{IRRE}
\end{equation}
In our case, $\mathbf{a},~\mathbf{b} = \mathbf{F},~\mathbf{I},~\mathbf{J},~\mathbf{L},~\mathbf{S}$ 
where $\mathbf{F} = \mathbf{I} + \mathbf{J}$ is the total angular momentum of the system and 
$\mathbf{J} = \mathbf{L} + \mathbf{S}$ is the total angular momentum of $\Omega^{-}$. Recall that 
$[a^{\alpha},a^{\beta}] = i\epsilon^{\alpha \beta \gamma} a^{\gamma}$ for
$\mathbf{a} = \mathbf{F},~\mathbf{I},~\mathbf{J},~\mathbf{L},~\mathbf{S}$,
$[F^{\alpha},a^{\beta}] = i\epsilon^{\alpha \beta \gamma} a^{\gamma}$ for 
$\mathbf{a} = \mathbf{I},~\mathbf{J},~\mathbf{L},~\mathbf{S}$,
$[J^{\alpha},a^{\beta}] = i\epsilon^{\alpha \beta \gamma} a^{\gamma}$ for 
$\mathbf{a} = \mathbf{L},~\mathbf{S}$, in other cases $[a^{\alpha},b^{\beta}] = 0$
for $\mathbf{a} \ne \mathbf{b}$. 

Let $\mathbf{a}+\mathbf{b}=\mathbf{c}$. Consider contractions of two tensors $\tau^{\alpha \beta}$:
\begin{eqnarray}
\nu(\mathbf{a},\mathbf{b},\mathbf{c}) &=& 
\tau^{\alpha \beta}(\mathbf{a},\mathbf{b})\tau^{\alpha \beta}(\mathbf{c},\mathbf{c}) \nonumber \\
&=& 4\mathbf{a \cdot c}\mathbf{b \cdot c} - \frac{4}{3} \mathbf{a \cdot b} \mathbf{c \cdot c}, \\
\nu(\mathbf{a},\mathbf{b}) &=& \tau^{\alpha \beta}(\mathbf{a},\mathbf{a})\tau^{\alpha \beta}(\mathbf{b},\mathbf{b}) \nonumber \\
&=& 4\mathbf{a \cdot b}\mathbf{a \cdot b} + 2\mathbf{a \cdot b} - \frac{4}{3} \mathbf{a \cdot a} \mathbf{b \cdot b},\\
\nu(\mathbf{a},\mathbf{c}) &=& \tau^{\alpha \beta}(\mathbf{a},\mathbf{a})\tau^{\alpha \beta}(\mathbf{c},\mathbf{c}) \nonumber \\
&=& 4\mathbf{a \cdot c}\mathbf{a \cdot c} - 2\mathbf{a \cdot c} - \frac{4}{3} \mathbf{a \cdot a} \mathbf{c \cdot c},\\
\nu(\mathbf{c}) &=& \tau^{\alpha \beta}(\mathbf{c},\mathbf{c})\tau^{\alpha \beta}(\mathbf{c},\mathbf{c}) \nonumber \\
&=& \frac{2}{3}c(c + 1)(2c - 1)(2c + 3),
\end{eqnarray}
where $\mathbf{c}\cdot \mathbf{c} = c(c + 1)$. 

Specifically, we define
\begin{equation}
\nu(\mathbf{n},\mathbf{L}) = \tau^{\alpha \beta}(\mathbf{n},\mathbf{n})\tau^{\alpha \beta}(\mathbf{L},\mathbf{L}) = - \frac{4L(L+1)}{3}. 
\end{equation}

Upon averaging over states with fixed $L$ (first line) or $J$ (other lines) 
one can write 
\begin{eqnarray}
\tau^{\alpha \beta}(\mathbf{n},\mathbf{n}) &=& \frac{\nu(\mathbf{n},\mathbf{L})}{\nu(\mathbf{L})} \tau^{\alpha \beta}(\mathbf{L},\mathbf{L}), \\
\tau^{\alpha \beta}(\mathbf{S},\mathbf{S}) &=& \frac{\nu(\mathbf{S},\mathbf{J})}{\nu(\mathbf{J})}\tau^{\alpha \beta}(\mathbf{J},\mathbf{J}), \\
\tau^{\alpha \beta}(\mathbf{S},\mathbf{L}) &=& \frac{ \nu(\mathbf{S},\mathbf{L},\mathbf{J})}{\nu(\mathbf{J})}\tau^{\alpha \beta}(\mathbf{J},\mathbf{J}), \\
\tau^{\alpha \beta}(\mathbf{L},\mathbf{L}) &=& \frac{ \nu(\mathbf{L},\mathbf{J})}{\nu(\mathbf{J})}\tau^{\alpha \beta}(\mathbf{J},\mathbf{J}).
\end{eqnarray}


The contraction of three functions $\tau^{\alpha \beta}$ entering Eq.(\ref{QQ}), averaged over a fixed $J$ state, gives
\begin{eqnarray}
&&\mu(\mathbf{S},\mathbf{L},\mathbf{J}) = \tau^{\alpha \beta}(\mathbf{S},\mathbf{S})\tau^{\beta \gamma}(\mathbf{L},\mathbf{L})\tau^{\gamma \alpha}(\mathbf{J},\mathbf{J}) \nonumber \\
&&= (2\mathbf{L} \cdot \mathbf{S} - \frac{1}{2})\nu(\mathbf{L},\mathbf{S},\mathbf{J}) - \frac{2}{3}S(S + 1)\nu(\mathbf{L},\mathbf{J})  - \frac{2}{3}L(L + 1)\nu(\mathbf{S},\mathbf{J}).
\end{eqnarray}

Using Ref.\cite{LLQM}, one gets  
\begin{equation}
<\frac{1}{r^3}> = \frac{2}{n^3L(L + 1)(2L + 1)a_{B}^3} 
\label{R3QUBE}
\end{equation}
and, for $P$-wave,
\begin{eqnarray}
R_{n1}^{\prime 2}(0) = \frac{4(n^2 - 1)}{9n^5a_{B}^5},
\label{R0}
\end{eqnarray}
where $a_{B} = 1/(\alpha Z m^{\prime})$ is the Bohr radius. 

The diagonal matrix elements of the interaction energies 
in the $J^{\prime}J$ basis can be found to be
\begin{eqnarray}
U_{IS} &=& \frac{3 \alpha Z g_{Z} g}{4mM}<\frac{1}{r^3}> 
\frac{\nu(\mathbf{n} , \mathbf{L}) }{\nu(\mathbf{L})}
\frac{ \mathbf{I} \cdot \mathbf{J}}{\mathbf{J} \cdot \mathbf{J}}
(\mathbf{J} \cdot \mathbf{L} \mathbf{L} \cdot \mathbf{S} - \frac{1}{3} \mathbf{L} \cdot \mathbf{L} \mathbf{J} \cdot \mathbf{S}), \label{AIS} \\
U_{Q_{Z}L} &=& - \frac{\alpha}{4}\frac{3Q_{Z}}{2I(2I - 1)}<\frac{1}{r^3}>
\frac{ 
\nu(\mathbf{n},\mathbf{L}) \nu(\mathbf{L},\mathbf{J}) \nu(\mathbf{I},\mathbf{J})
}
{
\nu(\mathbf{L}) \nu(\mathbf{J})
}, \label{AQL}\\
U_{LQ} &=& - \frac{\alpha Z}{4}\frac{3Q}{2S(2S - 1)}<\frac{1}{r^3}>
\frac{ 
\nu(\mathbf{n},\mathbf{L}) \nu(\mathbf{L},\mathbf{S})
}
{
\nu(\mathbf{L})
},
\end{eqnarray}
\begin{eqnarray}
U_{Q_{Z}S}^{cL} &=& \frac{\alpha g }{40 m m^{\prime}} \frac{3Q_{Z}}{2I(2I - 1)} \frac{\nu(\mathbf{I},\mathbf{J}) \nu(\mathbf{S},\mathbf{L},\mathbf{J})}{\nu(\mathbf{J})} R^{\prime 2}_{n1}(0), \label{AQS}\\
U_{IQ}^{cL}     &=& \frac{\alpha Z g_Z }{10 M m^{\prime}} \frac{3Q}{2S(2S - 1)} 
\frac{\mathbf{I} \cdot \mathbf{J}}{\mathbf{J} \cdot \mathbf{J}} \nonumber \\
&&\;\;\;\;\;\;\;\;\;\;\;\;\;\;\;\;\;\;\;\;\;\;\;\; \times \left(
 \mathbf{L} \cdot \mathbf{S} \mathbf{J} \cdot \mathbf{S} - \frac{1}{3}\mathbf{L} \cdot \mathbf{J} \mathbf{S} \cdot \mathbf{S}
\right)R^{\prime 2}_{n1}(0) \label{AIQ},\\
U_{Q_{Z}Q}^{c} &=& \frac{\alpha}{63} \frac{3Q_{Z}}{2I(2I - 1)} \frac{3Q}{2S(2S - 1)}
\frac{\nu(\mathbf{I},\mathbf{J})}{\nu(\mathbf{J})} \nonumber \\
&&\;\;\;\;\;\;\;\;\;\;\;\;\;\;\;\;\;\; \times \left(
\frac{7}{5} \nu(\mathbf{S},\mathbf{J})
- 3 \mu(\mathbf{S},\mathbf{L},\mathbf{J})
\right)R^{\prime 2}_{n1}(0). \label{AQQ}
\end{eqnarray}
Other matrix elements can be calculated using elementary tools.

The diagonal matrix element of $U_{Q_{Z}L}$ in the basis of fixed $\mathbf{I}+\mathbf{L}$ has the form
of Eq.(\ref{AQL}) with ${
\nu(\mathbf{L},\mathbf{J}) \nu(\mathbf{I},\mathbf{J})
}/
{
\nu(\mathbf{J})
}$ 
replaced by
$\nu(\mathbf{I},\mathbf{L})$ (cf. Ref. \cite{ABRA}, Chap. VI). 
In $J=1/2$ states of the $\Omega^{-}$ atoms, the diagonal element of the quadrupole - quadrupole interaction 
(\ref{AQQ}) vanishes, since $J=1/2$ states do not have quadrupole moments. The diagonal matrix element of 
$U_{Q_{Z}Q}^{c}$ for $J \neq 1/2$ is calculated in Refs. \cite{KARL1,KARL2}. Equation (\ref{AQQ}) is in agreement with Ref. \cite{KARL2}.

We use for calculations of the Clebsch-Gordan coefficients a code provided by Sierra \cite{CGCO}.
The diagonal matrix elements for the potentials entering $U^{[2]}$ and $U^{[4]}$ 
are in agreement with those calculated numerically. 

The contact interactions contribute to splitting of $L=1$ states and mixing and splitting of $L=0$ and $L=2$ states.

The numerical magnitudes of the contact $P$-wave interactions and the long-range interactions are compared 
by considering splitting of $2P$ energy levels of the $^{14}$N$\Omega ^{-}$ exotic atom. The nucleus 
$^{14}$N has spin $I=1$ and, respectively, magnetic and quadrupole moments. 


\begin{table}[tbp]
\scriptsize
\centering
\addtolength{\tabcolsep}{-3pt}
\caption{$J^{\prime}J$ matrix elements of fine and hyperfine interactions of order $(v/c)^2$ in
the $^{14}$N$\Omega ^{-}$ exotic atom for the $n=2$, $L=1$ state. 
$LS$ stands for the interaction energy $U_{LS}$ Eq.(\ref{LS}), 
$IL$ stands for the interaction energy $U_{IL}$ Eq.(\ref{IL}), and so on. 
Parameters used in the calculation: $\mu _{Z}=0.404$ n.m. \cite{PROC}, 
$\mu =-2.02$ n.m. \cite{WALL,DIEH}, $Q_{Z}=2.00$ fm$^{2}$ 
\cite{SCHI}, $Q=-2.8\times 10^{-2}$ fm$^{2}$ \cite{KRGI}. The energy is
given in keV. 
}
\label{tab:table1}
\vspace{2mm}

\begin{tabular}{|c|c||rrr|rrr|rrr|rrr|rrr|}
\hline
\multicolumn{2}{|c|}{U$^{[2]}$,~keV} & \multicolumn{3}{c|}{LS} & \multicolumn{3}{c|}{IL$\times 10$} & \multicolumn{3}{c|}{IS$\times 10$} & \multicolumn{3}{c|}{Q$_{Z}$L}
& \multicolumn{3}{c|}{LQ} \\ \hline
$F$ & $J$ & 1/2~  & 3/2~  & 5/2~ &  1/2~ &  3/2~ & 5/2~  &  1/2~ &  3/2~ & 5/2~  & 1/2~ & 3/2~ & 5/2~ & 1/2~ & 3/2~ & 5/2~ \\ \hline\hline
1/2 & 1/2 & -0.70 &  0.00 &      &  0.11 &  0.12 &       &  0.14 &  0.20 &       & 0.00 & 1.51 &  & -0.33 & 0.00 &  \\ 
    & 3/2 &  0.00 & -0.28 &      &  0.12 & -0.11 &       &  0.20 & -0.46 &       & 1.51 & -2.70 & & 0.00 & 0.26 &  \\ \hline
3/2 & 1/2 & -0.70 &  0.00 & 0.00 & -0.05 &  0.20 &  0.00 & -0.07 &  0.31 &  0.00 & 0.00 & -0.48 & 3.51 & -0.33 & 0.00 & 0.00 \\ 
    & 3/2 &  0.00 & -0.28 & 0.00 &  0.20 & -0.04 &  0.12 &  0.31 & -0.18 & -0.16 & -0.48 & 2.16 & -3.48 & 0.00 & 0.26 & 0.00 \\ 
    & 5/2 &  0.00 &  0.00 & 0.42 &  0.00 &  0.12 & -0.23 &  0.00 & -0.16 &  0.17 & 3.51 & -3.48 & 1.89 & 0.00 & 0.00 & -0.07 \\ \hline
5/2 & 3/2 &       & -0.28 & 0.00 &       &  0.07 &  0.15 &       &  0.28 & -0.20 &  & -0.54 & 1.86 &  & 0.26 & 0.00 \\ 
    & 5/2 &       &  0.00 & 0.42 &       &  0.15 & -0.65 &       & -0.20 &  0.05 &  & 1.86 & -2.16 &  & 0.00 & -0.07 \\ \hline
7/2 & 5/2 &       &       & 0.42 &       &       &  0.16 &       &       & -0.12 &  &  & 0.68 &  &  & -0.07 \\ 
\hline
\end{tabular}
\end{table}


The matrix elements of fine and hyperfine interactions are calculated
in the basis $J^{\prime }J$ at fixed $F$ for eight different contributions: spin-orbit $LS$ and $IL$, 
spin-spin $IS$, quadrupole-orbit $Q_{Z}L$ and $LQ$, quadrupole-spin $Q_{Z}S$ and 
$IQ$, and quadrupole-quadrupole $Q_{Z}Q$ interactions. The results for various terms 
entering the potential (\ref{U2}) are shown in Table \ref{tab:table1} and for the interaction (\ref{U4})
in Table \ref{tab:table2}. We restricted ourselves with estimates of the Larmor components of the 
quadrupole-spin interactions. 
In $2P$ state, the matrix elements of the contact interactions 
$<FJ^{\prime}|U^{[4]}|FJ>$ are suppressed as $(\alpha Z)^{2} \sim 3\times 10^{-3}$ with respect to 
the matrix elements of the long-range interactions $<FJ^{\prime}|U^{[2]}|FJ>$.

\begin{table}[tbp]
\scriptsize
\centering
\addtolength{\tabcolsep}{-3pt}
\caption{$J^{\prime}J$ matrix elements of contact $P$-wave interactions of order $(v/c)^4$ in 
the $^{14}$N$\Omega ^{-}$ exotic atom in the $n=2$, $L=1$ state. 
$IQ$ stands for the interaction energy $U_{IQ}$ Eq.(\ref{IQ}) and so on. 
Parameters and notations are the same as in Table \ref{tab:table1}.
}
\vspace{2mm}
\label{tab:table2}
\begin{tabular}{|c|c||rrr|rrr|rrr|}
\hline
\multicolumn{2}{|c|}{U$^{[4]}$,~keV} & \multicolumn{3}{c|}{IQ$\times 10^{4}$}
& \multicolumn{3}{c|}{Q$_{Z}$S$\times 10^2$} & \multicolumn{3}{c|}{Q$_{Z}$Q$\times 10^2$} \\ 
\hline
$F$ & $J$ & 1/2~  & 3/2~  & 5/2~ & 1/2~ & 3/2~ & 5/2~    &  1/2~ & 3/2~ & 5/2~ \\ 
\hline\hline
1/2 & 1/2 & -0.26 & -0.03 &      &  0.00 & -0.73 &       &  0.00 &  0.10 &  \\ 
    & 3/2 & -0.03 & -0.21 &      & -0.73 &  0.75 &       &  0.10 & -1.02 &  \\ \hline
3/2 & 1/2 &  0.13 & -0.05 & 0.00 &  0.00 &  0.23 & -0.49 &  0.00 & -0.03 & -0.96 \\ 
    & 3/2 & -0.05 & -0.08 & 0.09 &  0.23 & -0.60 & -0.24 & -0.03 &  0.81 & -0.37 \\ 
    & 5/2 &  0.00 &  0.09 & 0.11 & -0.49 & -0.24 &  0.78 & -0.96 & -0.37 & -0.07 \\ 
\hline
5/2 & 3/2 &       &  0.12 & 0.11 &       &  0.15 &  0.13 &       & -0.20 & 0.20 \\ 
    & 5/2 &       &  0.11 & 0.03 &       &  0.13 & -0.90 &       &  0.20 & 0.08 \\ \hline
7/2 & 5/2 &       &       &-0.08 &       &       &  0.28 &       &       &-0.03 \\ \hline
\end{tabular}
\end{table}

Our estimate of the contact $P$-wave quadrupole-quadrupole splitting in $^{14}$N$\Omega ^{-}$ 
is two orders of magnitude smaller than the estimate reported in Ref. \cite{KARL2}. 
The charge radius $r^2_{\Omega}$ included into the estimate of Ref. \cite{KARL2} can increase 
the hyperfine splitting, since $Q_{\Omega}$ has the smallness $\sim v/c$ or even 
$\sim (v/c)^3$ as compared to the proton and $\Omega^{-}$ charge radii, as discussed in Sect. 2. 
In Ref. \cite{KARL2}, it is assumed that $r^2_{\Omega}$ has the same magnitude as $Q_{\Omega}$, 
so the reason for the discrepancy is unclear. 
\footnote{This manuscript was in press when the authors were informed that G. Karl and V. A. Novikov 
revised their estimate of the hyperfine splitting. Their new estimate is in agreement with our 
[V. A. Novikov, private communication].}

The spin-orbit interaction is not dominant, probably except for $F=1/2$, 
so the total $\Omega ^{-}$ angular momentum $J$ does not provide diagonal basis. 
For $F=3/2,~5/3$, one has to diagonalize the energy operator in the space of
admissible $J$. For $F=7/2$ we have a $1\times 1$ matrix, so the values
given in Tables \ref{tab:table1} and \ref{tab:table2} for $F=7/2$ are the 
energy levels shifts. The effect of the contact interactions 
is comparable with the uncertainty in the experimental value of the $^{14}$N 
quadrupole moment, being two orders of the magnitude lower than the
quadrupole-orbit interaction.

\section{Natural widths and contact $P$-wave interactions in $\mu$-meson exotic atoms}

Sections 3.1 - 3.4 describe interactions of nuclei and particles with arbitrary masses and spins. 
These results can be applied to $\mu$-meson exotic atoms.
In heavy nuclei, the Bohr radius in $\mu$-meson exotic atoms is smaller than the nuclear radii and the problem 
is relativistic in addition. We discuss therefore the finite volume effects and the relativistic effects 
affecting the natural widths and the contact $P$-wave interactions.

\subsection{Natural widths of $\mu$-meson exotic atoms in $2P$ states}

Due to the dipole $2P-1S$ transition, the width of the $2P$ level equals
\begin{equation}
\Gamma_{2P}^{em} = \frac{4\alpha \omega_{fi}^3}{3} |\mathbf{x}_{fi}|^2 
                 = \left(\frac{2}{3}\right)^{8}\alpha (\alpha Z)^{4} \frac{m^{\prime 3}}{m^{\prime \prime 2}}, 
\label{2PNR}
\end{equation}
where $f=1S$, $i=2P$, $\omega_{fi} = E_{f} - E_{i}$, and
\begin{equation}
\frac{1}{m^{\prime \prime}} = \frac{Z}{M} + \frac{1}{m}. 
\end{equation}
It should be compared to the magnitude of the potential
$U_{Q_{Z}S} \sim \frac{1}{300}Q_{Z}m^{\prime 3}\alpha(\alpha Z)^{5}$. 
Condition $\Gamma_{fi}^{em} \ll U_{Q_{Z}S}$ gives roughly $Q_{Z} \gg 10/(\alpha Z m^{\prime 2}) \sim 6000/Z$ fm$^2$. The highest electric quadrupole moments of nuclei are about 500 fm$^2$, so it would make sense to check high-$Z$ nuclei.

The electromagnetic current has the form $\mathbf{j} = -eZ\mathbf{p}_{Z}/M + e\mathbf{p}/m$, 
where $\mathbf{p}_{Z}$ is momentum of the nucleus and $\mathbf{p}$ is momentum of the muon. In the center-of-mass frame, $\mathbf{j} = e\mathbf{p}/m^{\prime \prime}$. The quantity $e\mathbf{p}/m^{\prime}$
represents the convection current, which is the component of the total current $e\mbox{\boldmath{$\alpha$}}$, $\mbox{\boldmath{$\alpha$}}$ is the Dirac matrix. The nucleus spin current is neglected.
We use expression $\mathbf{j} = em^{\prime} $\mbox{\boldmath{$\alpha$}}$ /m^{\prime \prime}$.  For transition current, one has $\mathbf{j}_{fi} = ie\omega_{fi}m^{\prime} \mathbf{x}_{fi}/m^{\prime \prime}$. The dipole transition matrix element $\mathbf{x}_{fi}$ is calculated using wave functions obtained from solution of the Dirac equation.

\subsection{Electric charge and quadrupole moment densities in nuclei and nuclear electrostatic field}

Let us consider electrostatic potential, $\Phi_{0}$, created by the uniformly distributed electric charge inside of 
a sphere of radius $R=1.2 A^{1/3}$ fm:
\begin{equation}
\Phi_{0} = \left\{
\begin{tabular}{rl}
$- \left(\frac{3}{2} - \frac{r^2}{2R^2}\right) \frac{e Z}{R}$, & $r<R$, \\
$- \frac{e Z}{r}, $ & $r \geq R$,
\end{tabular}
\right.
\label{FINI0}
\end{equation}
so that we have for the electrostatic field
\begin{equation}
E^{\alpha}_{0} = \left\{ 
\begin{tabular}{rl}
$- \frac{r}{R} \frac{e Z}{R^2}n^{\alpha}$, & $r<R$, \\
& \\
$- \frac{e Z}{r^{2}}n^{\alpha}, $ & $r \geq R$.
\end{tabular}
\right.
\label{FINI1}
\end{equation}
The charge density is given by
\begin{equation}
4\pi \rho_{0} = \mathrm{div} \mathbf{E}_{0} = \left\{ 
\begin{tabular}{rl}
$- eZ \frac{3}{R^3}$, & $r<R$, \\
$0, $ & $r \geq R$,
\end{tabular}
\right.
\label{FINI2}
\end{equation}
and normalized to
\begin{equation}
\int \rho_{0} dV = -eZ >0.
\end{equation}

The quadrupole component of the electrostatic potential has the form
\begin{equation}
\Phi_{2} = \left\{
\begin{tabular}{rl}
$\frac{5r^2}{2R^5} \ln(\frac{R_{1}}{r}) Q^{\alpha \beta}_{Z} n^{\alpha} n^{\beta}$, & $r<R$, \\
& \\
$\frac{1   }{2r^3} Q^{\alpha \beta}_{Z} n^{\alpha} n^{\beta},$ & $r \geq R$,
\end{tabular}
\right.
\label{QUAD1}
\end{equation}
where $R_1 = R e^{1/5}$. The radial dependence of $\Phi_{2}$ corresponds to the uniform radial distribution of the quadrupole component of the electric charge. 

The quadrupole electrostatic field $\mathbf{E}_{2}$ can be decomposed to the sum of $L=1$ and $L=3$ components, the former constitutes the analogue of the delta-function component discussed in Sect. 3.4:
\begin{eqnarray}
E_{21}^{\alpha} &=&  \left\{
\begin{tabular}{rl}
$- \frac{5r}{R^5} \ln(\frac{R}{r}) Q^{\alpha \beta}_{Z} n^{\beta},$ & $r<R$, \\
& \\
$0$, & $r \geq R$, 
\end{tabular}
\right.
\label{E111} \\
E_{23}^{\alpha} &=&  \left\{
\begin{tabular}{rl}
$ \frac{5r}{2R^5} Q^{\beta \gamma}_{Z} T^{\alpha \beta \gamma}_{3},$ & $r<R$, \\
& \\
$ \frac{5}{2r^4}  Q^{\beta \gamma}_{Z} T^{\alpha \beta \gamma}_{3},$ & $r \geq R$, 
\end{tabular}
\right.
\label{E333}
\end{eqnarray}
where
\[
T^{\alpha \beta \gamma}_{3} = n^{\alpha} n^{\beta} n^{\gamma} - \frac{1}{5}
\left(\delta^{\alpha \beta }n^{\gamma} + \delta^{\beta \gamma}n^{\alpha} + \delta^{\gamma \alpha}n^{\beta}\right).
\]
The fact that the $L=1$ component is localized inside the nucleus indicates that we deal with a contact interaction.
The divergence of the $L=3$ component of the quadrupole electrostatic field contributes to the charge density at $r<R$ and the quadrupole moment also, so that we have
\begin{eqnarray}
4\pi \rho_{21} = \mathrm{div} \mathbf{E}_{21} &=&  \frac{5 }{R^{5} }Q^{\alpha \beta}_{Z}n^{\alpha}n^{\beta}, \;\;\; r<R,\label{QUAD5} \\
4\pi \rho_{23} = \mathrm{div} \mathbf{E}_{23} &=&  \frac{15}{2R^{5}}Q^{\alpha \beta}_{Z}n^{\alpha}n^{\beta},\;\;\; r<R,  \label{QUAD6}
\end{eqnarray}
and $\rho_{21} = \rho_{23} = 0$ for $r\geq R$. The normalization of $\Phi_{2}$ is chosen to satisfy
\begin{equation}
\int (3x^{\alpha}x^{\beta} - r^{2} \delta^{\alpha \beta}) \rho_{2} dV   = Q^{\alpha \beta}_{Z}.
\label{QUAD3}
\end{equation}
where $\rho_{2} = \rho_{21} + \rho_{23}$.

\subsection{Point size nucleus and nonrelativistic approximation}

The nonrelativistic reduction of the Dirac equation in external electrostatic field gives
\begin{equation}
\Delta U = - \frac{e(g-1)}{8m^2} \left( 
[\mathbf{E}\times \mathbf{p}] \mbox{\boldmath{$\sigma$}} 
 - [\mathbf{p} \times \mathbf{E}] \mbox{\boldmath{$\sigma$}} + \mathrm{div}\mathbf{E} 
\right).
\end{equation}
(for $g=2$ see Ref. \cite{BJDR}). Substituting in this expression $\mathbf{E}_{2} = \mathbf{E}_{21} + \mathbf{E}_{23}$ and integrating the short-range part of the matrix element of $\Delta U$ at $|\mathbf{x}|<R$ one gets, in the limit of $R \to 0$, 
\begin{equation}
\Delta U^{c} = \frac{e(g-1)}{80m^2} Q_{Z}^{\alpha \beta} 
(5 \tau^{\alpha \beta}(\mathbf{L},\mathbf{L}) - \tau^{\alpha \beta}(\mathbf{L},\mathbf{S})) R_{n1}^{\prime 2}(0).
\label{NREL}
\end{equation}
The numerical coefficients in front of the spin operators are distinct from those in Eq.(\ref{QSFULL}), since Eq.(\ref{NREL}) includes the effect of the $L=3$ component of $\mathbf{E}_{2}$ and the Darwin term $\sim \mathrm{div}\mathbf{E}$.

The radial part of the long-range component $|\mathbf{x}|>R$ of the quadrupole interaction diverges logarithmically at $R \to 0$. The spin matrix element, however, vanishes in $L=1$ state, so
$\Delta U^{long-range} = 0$.

The Dirac equation contains the contact $P$-wave interactions. Relativistic wave equations for high-spin particles 
\cite{KRIV,NIED}, obviously, contain the contact $P$-wave interactions also. The Darwin term for $S=3/2$ has the form
\begin{equation}
\Delta U_{D} = - \frac{e(12g-7)}{24m^2} \mathrm{div}\mathbf{E}.
\label{DARW}
\end{equation}
The coefficient can be restored using Eq.(9b) of Ref. \cite{GIAN}.


\begin{table}[tbp]
\scriptsize
\centering
\addtolength{\tabcolsep}{-1pt}
\caption{
The magnitudes of quadrupole-orbit long-range and quadrupole-spin contact $P$-wave interactions and natural decay widths 
of $2P$ states of muonic atoms formed with several spin $I \geq 1$ low- and high-$Z$ nuclei. The experimental values of the nuclear 
electric quadrupole moments $Q_{Z}$ are taken from Ref. \cite{RAGH}, errors are not displayed.
$U^{[2]}_{\max}$ is the maximum of the absolute value over $F$ of the diagonal matrix elements $<FJ|U_{Q_{Z}L}|FJ>$ for $J = 3/2$ and $n=2$, $U^{[4]}_{\max}$ is defined similarly for $U_{Q_{Z}S}^{c}$. The effects of relativity and the finite volume of nuclei are included. $\Gamma_{2P}^{em}$ is the radiation width of the $2P_{3/2}-1S_{1/2}$ transition.
}
\label{tab:table91}
\begin{tabular}{|c|ccccccc|} \hline
 Nuclei               & $^{2}$H &$^{6}$Li & $^{7}$Li & $^{9}$Be   & $^{10}$B & $^{11}$B &$^{14}$N \\ \hline \hline
 $I$                  & 1       &1        & 3/2      & 3/2        & 3        & 3/2      &1        \\ \hline
 $Q_{Z}$   [fm$^2$]   & 0.29    &-0.08    & -4.06    & 5.3        & 8.47     & 4.07     &2.00	  \\ \hline
$U^{[2]}_{\max}$ [eV]&$4.4\times 10^{-4}$&$3.7\times 10^{-3}$&$1.9\times 10^{-1}$&$5.8\times 10^{-1}$&$8.8\times 10^{-1}$&$8.8\times 10^{-1}$&$1.2$  \\ \hline
$U^{[4]}_{\max}$ [eV]&$5.8\times 10^{-9}$&$4.2\times 10^{-7}$&$2.1\times 10^{-5}$&$1.2\times 10^{-4}$&$2.7\times 10^{-4}$&$2.8\times 10^{-4}$&$7.2\times 10^{-4}$\\ \hline
$\Gamma_{2P}^{em}$ [eV]   &$8.8\times 10^{-5}$&$7.3\times 10^{-3}$&$7.2\times 10^{-3}$&$2.3\times 10^{-2}$&$5.8\times 10^{-2}$&$5.8\times 10^{-2}$& $2.2\times 10^{-1}$\\ \hline
\hline
 Nuclei               & $^{181}$Ta &$^{185}$Re & $^{190}$Ir & $^{193}$Ir & $^{197}$Au & $^{235}$U & $^{253}$Es   \\ \hline \hline
 $I$                  & 7/2        & 5/2       & 4          & 3/2        & 3/2        & 7/2       & 7/2          \\ \hline
 $Q_{Z}$  [fm$^2$]    & 317        & 218       & 285        & 75.1       & 54.7       &  493      & 670          \\ \hline
$U^{[2]}_{\max}$ [eV]&$5.9\times 10^{4}$&$5.5\times 10^{4}$&$5.3\times 10^{4}$&$3.5\times 10^{4}$&$2.7\times 10^{4}$&$1.2\times 10^{5}$&$1.7\times 10^{5}$\\ \hline
$U^{[4]}_{\max}$ [eV]&$1.7\times 10^{3}$&$1.7\times 10^{3}$&$1.7\times 10^{3}$&$1.1\times 10^{3}$&$8.6\times 10^{2}$&$4.6\times 10^{3}$&$7.0\times 10^{3}$\\ \hline
$\Gamma_{2P}^{em}$ [eV]   &$7.2\times 10^{2}$&$7.7\times 10^{2}$&$8.4\times 10^{2}$&$8.3\times 10^{2}$&$9.0\times 10^{2}$&$1.4\times 10^{3}$&$1.9\times 10^{3}$\\ \hline
\end{tabular}
\end{table}


\subsection{Finite size nucleus and relativistic approximation}

The nonrelativistic limit of the Dirac equation for $g=2$ and $M \to \infty$ results to the quadrupole-spin 
contact $P$-wave interaction originating from the lower components of the Dirac bispinors. The lower components 
produce, however, other interactions too. We restrict ourselves with evaluation of the Larmor and Thomas contact $P$-wave interactions corresponding to $\mathbf{E}_{21}$.

In order to discriminate the contact interactions, we use the Gordon's decomposition of the electromagnetic current: 
\begin{equation}
j_{\mu} = j_{\mu}^{conv} + j_{\mu}^{spin} = \frac{e}{2m}\bar{\psi}\left((i \stackrel{\leftrightarrow}\partial_{\mu} - 2eA_{\mu}) + \frac{g}{2}\sigma_{\mu \nu} (\stackrel{\leftarrow}\partial + \stackrel{\rightarrow}\partial)_{\nu}\right)\psi,
\end{equation}
where $A_{\mu} = (\Phi_{0},\mathbf{0})$, $\Phi_{0}$ is given by Eq.(\ref{FINI0}). The time-like component of the spin current interacts with the quadrupole electrostatic potential:
\begin{equation}
U^{L}_{Q_{Z}S} = \int \Phi_{2} j_{0}^{spin} dV.
\label{QQQQ}
\end{equation}

The interaction energy (\ref{QQQQ}) corresponding to $E_{21}^{\alpha}$ constitutes the relativistic counterpart of the Larmor contact $P$-wave interaction discussed in Sect. 3.4. The relativistic extension of Eq.(\ref{QS}), which 
takes the finite size of the nucleus into account, can be written as follows:
\begin{eqnarray*}
U^{cL}_{Q_{Z}S} &=& \frac{\alpha g}{40m m^{\prime}} \frac{3Q_{Z}}{2I(2I - 1)} 
\tau^{\alpha \beta}(\mathbf{I},\mathbf{I})\tau^{\alpha \beta}(\mathbf{S},\mathbf{L}) \\
&&\;\;\;\;\;\;\;\;\;\;\;\;\;\;\;\;\;\;\;\;\;\;\;\; \times \int_{0}^{R}r^2 dr \frac{25}{R^{5}}\ln(\frac{R}{r})\frac{2m^{\prime}}{\varepsilon + m^{\prime} - V}f_{nJL}^{2}(r),
\label{FFFF}
\end{eqnarray*}
where $f_{nJL}(r)$ is the upper radial component of the Dirac wave function in the potential $V=e\Phi_{0}$, with the normalization conventions of Ref. \cite{LALI}, and $\varepsilon$ is the energy of the $nJL$ level. In the limit of $Z \to 0$ and $R \to 0$, $f_{nJL}(r) \to R_{nL}(r)$ and we recover Eq.(\ref{QS}). In order to arrive at Eq.(\ref{FFFF}), we drop a term proportional to $\mathrm{div} \mathbf{E}_{21}$ whose physical origin is attributed to the muon Zitterbewegung. It contributes to hyperfine structure of the $P$-wave levels and is included into Eq.(\ref{NREL}).

The Larmor component of the interaction is localized completely inside of the nucleus. 
To get the correct isotope dependence in the limit of $Z \rightarrow 0$, one has to use the mass $m$ in the electromagnetic current and the reduced mass $m^{\prime}$ in the Dirac equation. 

The interaction of the convection current $j_{\mu}^{conv}$ with the quadrupole electrostatic 
potential generates the long-range quadrupole-orbit interaction and the Thomas component of 
the quadrupole-spin interaction. The latter comes from the lower components of the bispinors.
The relativistic extension of the Thomas component of the quadrupole-spin contact interaction has the form
\begin{eqnarray*}
U^{cT}_{Q_{Z}S} &=& - \frac{\alpha}{40m^{2}} \frac{3Q_{Z}}{2I(2I - 1)} 
\tau^{\alpha \beta}(\mathbf{I},\mathbf{I})\tau^{\alpha \beta}(\mathbf{S},\mathbf{L}) \\
&&\;\;\;\;\;\;\;\;\;\;\;\;\;\;\;\;\;\;\;\;\;\;\;\; \times \int_{0}^{R}r^2 dr \frac{25}{R^{5}}\ln(\frac{R}{r})\frac{4m^{\prime}(\varepsilon - V)}{(\varepsilon + m^{\prime} - V)^{2}}f_{nJL}^{2}(r).
\label{GGGG}
\end{eqnarray*}
Correct isotope dependence in the limit of $Z \rightarrow 0$ is reproduced by using the $m^{\prime}$ in the Dirac wave functions. The off-diagonal matrix elements receive an additional dependence on quantum numbers $JL$ from the radial integrals.

In fixed-$J$ multiplets at $Z \to 0$ and $R \to 0$, the ratio between strengths of the Larmor component of the contact $P$-wave quadrupole-spin energy and the long-range quadrupole-orbit energy (i.e. between the right-hand sides of Eqs.(B.20) and (B.18)) equals 
$(\alpha Z)^{2}/4$. Due to the relativistic and finite volume corrections, the width $\Gamma_{2P}^{em}$ in $^{235}$U receives an additional factor of $0.22$ as compared to the nonrelativistic formula (\ref{2PNR}), 
the average value of $1/r^{3}$ entering the quadrupole-orbit interaction receives a factor of $0.42$, the Larmor component 
of the contact $P$-wave interaction is suppressed by a factor of $0.15$, and the Thomas component of the contact $P$-wave interaction is suppressed by a factor of $0.16$. The numbers are given for $2P_{3/2}$ state. 

Results reported in Table \ref{tab:table91} give an idea about the magnitudes of the natural widths and 
the contact $P$-wave interactions in muonic atoms formed with low- and high-$Z$ nuclei.

\section{Spin-1/2 and spin-3/2 relativistic spinors and $SU_{6}$ 
relations for octet and decuplet coupling constants}
\setcounter{equation}{0}

The relativistic spinors of spin-1/2 and spin-3/2 particles are normalized by
\begin{eqnarray}
\bar{u}(p,s)u(p,s) &=& 1, \\
-\bar{u}_{\mu }(p,s)u_{\mu }(p,s) &=& 1.
\end{eqnarray}

In the rest frame, these spinors have the form 
\begin{eqnarray}
u &=& \left( 
\begin{array}{l}
\varphi \\ 
0
\end{array}
\right) ,\\
u^{0}&=&\left( 
\begin{array}{l}
0 \\ 
0
\end{array}
\right) ,\;\;\;\mathbf{u}=\left( 
\begin{array}{l}
\mbox{\boldmath{$\xi$}} \\ 
0
\end{array}
\right),
\end{eqnarray}
where $\varphi$ is the Pauli spinor and $\mbox{\boldmath{$\xi$}}$ is the rest-frame spin-vector obeying the
condition $\mbox{\boldmath{$\sigma$}} \cdot \mbox{\boldmath{$\xi$}} = 0$ needed to eliminate the spin-1/2 component from $\mbox{\boldmath{$\xi$}}$:
\begin{equation}
\mbox{\boldmath{$\xi$}}(m) = \sum_{\mu} C^{\frac{3}{2} m}_{\frac{1}{2} \mu 1 m - \mu}\xi(\mu) \mathbf{e}^{(m - \mu)}. 
\label{raritaspinor}
\end{equation}
Here, the vector $\mathbf{e}^{(m)}$ with spin projection $m$ is defined by
\begin{equation}
\mathbf{e}^{(m)}  = \frac{i}{\sqrt{2}} \mbox{\boldmath{$\sigma$}}_{\beta }^{\alpha }\xi_{\alpha }^{\beta}(m),
\end{equation}
where $\xi ^{11}(+1)  = \xi ^{22}(-1)=\sqrt{2}\xi ^{12}(0)=\sqrt{2}\xi ^{21}(0)=1$, 
other components of spin-tensor $\xi ^{\alpha \beta}(m)$ vanish, $\xi_{\alpha }^{\beta}(m) = C_{\alpha \gamma } \xi ^{\beta \gamma}(m)$, 
and $C_{\alpha \beta } = i(\sigma ^{2})_{\beta }^{\alpha}$.

The spinors and spin-vectors with fixed spin projections are normalized conventionally:
\begin{eqnarray}
\sum_{\alpha}  \varphi^{*}_{\alpha} (m^{\prime}) \varphi^{\alpha} (m) &=& \delta_{m^{\prime} m}, \\
\sum_{\alpha i}\xi^{i *}_{  \alpha} (m^{\prime}) \xi^{  \alpha i} (m) &=& \delta_{m^{\prime} m}.
\end{eqnarray}
The completeness conditions have the form
\begin{eqnarray}
\frac{1}{2s+1} \sum_{m}\varphi^{\alpha} (m) \varphi^{*}_{\beta}(m) &=& \frac{1}{2}\delta^{\alpha}_{\beta}, \\
\frac{1}{2s+1} \sum_{m}\xi^{\alpha i}   (m) \xi^{j *}_{\beta } (m)&=& \frac{1}{6}
\left(\delta^{\alpha}_{ \beta}\delta^{ij} - \frac{i}{2} \epsilon^{ijk}(\sigma^{k})^{\alpha}_{ \beta}\right).
\end{eqnarray}

Applying the boost transformation to the spinor indices of $u$ and $u_{\mu}$ (see e.g. \cite{BJDR}, Chap. 3), 
and additionally, to the vector indices of $u_{\mu}$ using the matrix $L(\mathbf{v})$ of Eq.(\ref{LORE}), one gets the 
relativistic spinors:
\begin{eqnarray}
u(p,s) &=&\sqrt{\frac{E+m}{2m}}\left( 
\begin{array}{l}
\varphi \\ 
\frac{1}{E+m}(\mbox{\boldmath{$\sigma$}} \mathbf{p}) \varphi
\end{array}
\right) , \label{DI} \\
u^{0}(p,s) &=&\sqrt{\frac{E+m}{2m}}\left( 
\begin{array}{l}
\frac{1}{m}(\mathbf{p} \mbox{\boldmath{$\xi$}}) \\ 
\frac{1}{m(E+m)}( \mbox{\boldmath{$\sigma$}} \mathbf{p})(\mathbf{p} \mbox{\boldmath{$\xi$}})
\end{array}
\right) ,\nonumber \\
\mathbf{u}(p,s)&=&\sqrt{\frac{E+m}{2m}}\left( 
\begin{array}{l}
\mbox{\boldmath{$\xi$}} + \frac{1}{m(E+m)} \mathbf{p} (\mathbf{p} \mbox{\boldmath{$\xi$}} ) \\ 
\frac{1}{E+m}( \mbox{\boldmath{$\sigma$}} \mathbf{ p} ) (\mbox{\boldmath{$\xi$}} 
+ \frac{1}{m(E+m)} \mathbf{p} (\mathbf{p}\mbox{\boldmath{$\xi$}} ))
\end{array}
\right). \label{RS}
\end{eqnarray}

One can check that $u_{\mu}$ obeys $p_{\mu}u_{\mu}(p,s)=0$ 
and $\gamma_{\mu}u_{\mu}(p,s)=0$.

Expressions (\ref{DI}) and (\ref{RS}) can be used to get the nonrelativistic reduction 
Eqs.(\ref{EV1}) - (\ref{sJ2}) of the vertices $N\Lambda K$ and $\Omega \Xi K$.

We use the $SU_{6}$ symmetry to fix sign of the interference term. Firstly, we get 
relations for the isovector pseudoscalar coupling constants $g_{NN\pi}$ and $g_{\Delta N \pi}$ and
the isovector vector coupling constants $g_{NN\rho}$, $f_{NN\rho}$ and $f^{[1]}_{\Delta N \rho}$.
The relations for strange baryons can be obtained afterwards like in Sect. 4.3 
with the use of the $SU_{3}$ symmetry.

The source of the pion field is divergence of the isovector axial vector 
current
\begin{equation}
J_{P}(0)=i\lambda _{P}q^{\alpha }\sum_{i}\sigma _{i}^{\alpha }\tau _{i}^{3},
\label{JP}
\end{equation}
where the summation extends to quarks, $\lambda _{P}$ is an unknown real constant, 
and $q^{\alpha }$ is the pion momentum.

In the $\Delta \to N \rho$ transition, the magnetic component is dominant. The source 
of the $\rho$ mesons field of the magnetic type is rotor of the isovector axial 
vector current
\begin{equation}
J_{V}^{\alpha }(0)=i\lambda _{V}\epsilon ^{\alpha \beta \gamma }q^{\beta}\sum_{i}\sigma _{i}^{\gamma }\tau _{i}^{3},
\label{JV}
\end{equation} 
where the summation extends to quarks, $\lambda _{V}$ is an unknown real constant, 
and $q^{\alpha }$ is the $\rho$-meson momentum.

The matrix elements of the operator $\sum_{i}\sigma _{i}^{\alpha }\tau _{i}^{3}$ over the proton and $\Delta^{+}$ quark 
wave functions with the spin projections $+1/2$ are expressed in terms of the matrix elements of the corresponding baryons wave functions:
\begin{eqnarray}
<p,+\frac{1}{2}|\sum_{i}\sigma _{i}^{\alpha }\tau _{i}^{3}|p,+\frac{1}{2}> &=& \frac{5}{3}\varphi ^{+}\sigma^{\alpha }\varphi, \label{pp} \\
<p,+\frac{1}{2}|\sum_{i}\sigma _{i}^{\alpha }\tau _{i}^{3}|\Delta ^{+},+\frac{1}{2}> &=&
- \frac{4}{\sqrt{3}}\varphi ^{+} \epsilon^{\alpha \beta \gamma} \sigma^{\beta} \xi ^{\gamma} =
\frac{4i}{\sqrt{3}}\varphi ^{+} \xi ^{\alpha}. \label{pd}
\end{eqnarray}
In the right-hand sides, $\sigma^{\alpha }$ act on the baryon spin indices. 
In deriving these equations, we used the $\Delta$ and proton wave functions 
constructed by composition like in Eq.(\ref{raritaspinor}) of the spin-1/2 d-quark and spin-1 uu-diquark wave functions.

Comparison with the nonrelativistic matrix elements of $J_P(0)$ and $J^{\alpha}_V(0)$ of Sect. 4 gives
${g_{ N     N \pi}}/{(2 m_N)} = - \frac{5}{3} \lambda_P$,
${g_{\Delta N \pi}}/{m_N}   = 2\sqrt{2}   \lambda_P$
and
${(g_{NN\rho }+f_{NN\rho })}$ $/{(2m_{N})} = - \frac{5}{3}\lambda _{V}$,
${m_{\Delta }f^{[1]}_{\Delta N\rho }}/{m_{N}^{2}} = 2 \sqrt{2} \lambda _{V}$. In order to pass from 
$p \pi^0$ channel to $p \pi^+$ channel, we take into account the factor $\sqrt{3/2}$; similarly for 
$p\rho$ channels.

Finally, we obtain
\begin{eqnarray}
g_{\Delta N\pi } &=& - \frac{3\sqrt{2}}{5} g_{NN\pi }, \\
f^{[1]}_{\Delta N\rho }&=& - \frac{3\sqrt{2}}{5}\frac{m_{N}}{m_{\Delta }}(g_{NN\rho }+f_{NN\rho}).
\end{eqnarray}

The value of the $\Delta N\pi$ coupling constant is slightly away from the empirical value 
(\ref{OXiK}) for $g_{N N\pi }/\sqrt{4\pi} = 3.67$. The $\Delta N\rho$ coupling constant is found to be
$f^{[1]}_{\Delta N\rho }/\sqrt{4\pi}= - 2.82$ for $g_{NN\rho }/\sqrt{4\pi} = 0.84$ and 
$f_{NN\rho }/\sqrt{4\pi} = 3.53$ \cite{STOC99}, model NCS97a, in the excellent agreement with Eq.(\ref{DELTA1}). 

The relative phases of the octet and decuplet coupling constants are thereby fixed. Coming back to Eq.(\ref{R13}), 
we observe that the interference term $\mathcal{R}_{13}$ in the $\Lambda \Xi$ decay channel is positive. 
The remaining ambiguities affect phase of 
the total amplitudes, but not the interference. 

Assuming the $SU_{6}$ symmetry holds, the pseudoscalar meson exchnage and the magnetic vector meson exchange 
interfere in $S$-wave constructively:

Consider first the nonstrange sector. In the nonrelativistic approximation, the PWBA amplitude of reaction $AB \to CD$ 
due to $\pi^0$ and $\rho^0$ exchanges can be written as follows
\begin{eqnarray}
\mathfrak{M} &\sim& <C|J_P(0)|A><D|J_P(0)|B> \frac{1}{q^2 - m_{P}^2} \nonumber \\
             &+&    <C|J_V^{\alpha}(0)|A><D|J_V^{\alpha}(0)|B> \frac{1}{q^2 - m_{V}^2}.
\end{eqnarray}
Using Eqs.(\ref{JP}) and (\ref{JV}) and averaging the amplitude over the directions of the momentum transferred $\mathbf{q}$,
one arrives at
\begin{equation}
\mathfrak{M} \sim  <C|\sum_{i}\sigma _{i}^{\alpha }\tau _{i}^{3}|A> 
                   <D|\sum_{i}\sigma _{i}^{\alpha }\tau _{i}^{3}|B>  
                  (             \frac{\lambda_{P}^2 m_{P}^{*2}}{\mathbf{q}^2 + m_{P}^{*2}} 
                   + \frac{2}{3}\frac{\lambda_{V}^2 m_{V}^{*2}}{\mathbf{q}^2 + m_{V}^{*2}}).
\end{equation}
As we mentioned, $\lambda_P$ and $\lambda_V$ are real constants (as a consequence of hermiticity of the currents), 
and so the two terms in brackets both are positive. 

If we replace $\tau _{i}^{3}$ by $\mbox{\boldmath{$\tau$}} _{i}$ and $\mbox{\boldmath{$\tau$}} _{i}$ by $U$- and 
$V$-spin generators of the $SU_{3}$ group, the statement on the constructive interference extends further to 
the charged $\pi$- and $\rho$-mesons and kaons.

The pseudoscalar and magnetic vector coupling constants are therefore proportional to the same quark operator. If the 
ratio between $g_{NN\pi}$ and $g_{NN\rho} + f_{NN\rho}$ is taken positive, it remains positive for other members 
of the pseudoscalar and vector meson octets. The model NCS97a \cite{STOC99} fulfills such requirement for all 
coupling constants except for $\Xi \Xi M$ and $\Sigma N M$, whereas the model NCS97f \cite{STOC99} fulfills it without exceptions.
So, by following the model NCS97a we arrive at a destructive $S$-wave interference in the $\Sigma \Xi$ channel.
The model NCS97f predicts a constructive interference there. The models NCS97a and NCS97f both predict 
constructive interference in the $S$-wave dominant $\Lambda \Xi$ channel in agreement with $SU_{6}$.

\section{Angular part of gradient}
\setcounter{equation}{0}

In the momentum representation,
the angular part $\bigtriangledown ^{\alpha }$ of gradient is defined as 
operator $p\frac{\partial}{\partial \mathbf{p}}$ acting
on functions of unit vectors $\mathbf{n}=\mathbf{p}/p$  (see e.g. \cite{LALI}, Chap. VII). 
The knowledge of identities listed below allows to simplify the calculation 
of integrals entering Eqs.(\ref{I1}) - (\ref{I4}):
\begin{eqnarray}
L ^{\alpha } &=&i\epsilon ^{\alpha \beta \gamma }\bigtriangledown ^{\beta
}n^{\gamma },  \label{B17} \\
\lbrack \bigtriangledown ^{\alpha },\bigtriangledown ^{\beta }] &=& i\varepsilon^{\alpha \beta \gamma} L^{\gamma} = 
n^{\alpha} \bigtriangledown ^{\beta }-n^{\beta }\bigtriangledown ^{\alpha }, \label{B11}\\
\lbrack \bigtriangledown ^{\alpha },n^{\beta }] &=& \delta ^{\alpha \beta
}-n^{\alpha }n^{\beta },  \label{B12}\\
n^{\alpha }\bigtriangledown ^{\alpha } &=&0,  \label{B13} \\
\frac{\partial }{\partial p_{\alpha }} &=&n^{\alpha }\frac{\partial }{%
\partial p}+\frac{1}{p}\bigtriangledown ^{\alpha },  \label{B14} \\
\frac{\partial }{\partial p_{\alpha }}\frac{\partial }{\partial p_{\beta }}
&=&n^{\alpha }n^{\beta }\frac{\partial ^{2}}{\partial p^{2}}+(\delta
^{\alpha \beta }-n^{\alpha }n^{\beta }+n^{\beta }\bigtriangledown ^{\alpha
}+n^{\alpha }\bigtriangledown ^{\beta })\frac{1}{p}\frac{\partial }{\partial
p} \nonumber \\
&& \;\;\;\;\;\;\;\;\;\;\;\;\;\;\;\;\;\;\;\;\;\;\;\;+\frac{1}{p^{2}}(\bigtriangledown ^{\alpha }\bigtriangledown ^{\beta
}-n^{\alpha }\bigtriangledown ^{\beta }),  \label{B15} \\
\frac{\partial }{\partial p_{\alpha }}\frac{\partial }{\partial p_{\alpha }}
&=&\frac{\partial ^{2}}{\partial p^{2}}+\frac{2}{p}\frac{\partial }{\partial
p}+\frac{1}{p^{2}}\bigtriangledown ^{\alpha }\bigtriangledown ^{\alpha },  \label{B16} \\
\bigtriangledown ^{\alpha }\bigtriangledown ^{\alpha } &=&-L ^{\alpha
}L ^{\alpha }. 
\end{eqnarray}


The integrals of angular variables, entering Eqs.(\ref{I1}) - (\ref{I4}), 
obey the following properties:
\begin{eqnarray}
\int Y_{l^{\prime }m^{\prime }}^{*}Y_{lm}d\Omega _{\mathbf{n}} &=&\delta
_{l^{\prime }l}\delta _{m^{\prime }m}, \label{B1} \\
\int \left( \bigtriangledown ^{\alpha }Y_{l^{\prime }m^{\prime }}^{*}\right)
\left( \bigtriangledown ^{\alpha }Y_{lm}\right) d\Omega _{\mathbf{n}}
&=&l(l+1)\delta _{l^{\prime }l}\delta _{m^{\prime }m},  \label{B3}  \\
\int \left( L ^{\alpha }Y_{l^{\prime }m^{\prime }} ^{*} \right) \left( L
^{\alpha }Y_{lm}\right) d\Omega _{\mathbf{n}} &=&l(l+1)\delta _{l^{\prime
}l}\delta _{m^{\prime }m},  \label{B4} \\
\int \left( \bigtriangledown ^{\alpha } \bigtriangledown ^{\beta }Y_{l^{\prime }m^{\prime }}^{*}\right) 
\left( \bigtriangledown^{\alpha } \bigtriangledown ^{\beta }Y_{lm}\right) d\Omega _{\mathbf{n}} &=&l^2(l+1)^2 \delta _{l^{\prime
}l}\delta _{m^{\prime }m},  \label{B5} 
\end{eqnarray}
where $l$ is the orbital quantum number.
Equation (\ref{B1}) is the orthogonality condition for spherical harmonics $Y_{lm}(\mathbf{n})$, 
the next two equations give the normalization condition for the electric and magnetic 
spherical vectors.

\end{appendix}


\end{document}